\documentclass[aps,prl,twocolumn,superscriptaddress,longbibliography,floatfix,10pt]{revtex4-2}
\usepackage{bm,amsmath,amssymb}
\usepackage{graphicx,xcolor,overpic,float}
\usepackage{placeins}
\colorlet{BLUE}{blue}
\definecolor{sepTwentyPurple}{RGB}{145,35,180}

\colorlet{newpurple}{blue}
\colorlet{NEWPURPLE}{newpurple}

\colorlet{revisiongreen}{blue}

\colorlet{revisionorange}{blue}

\colorlet{revisionpolish}{blue}
\colorlet{REVISIONPOLISH}{revisionpolish}

\usepackage[hidelinks,breaklinks=true,hypertexnames=false]{hyperref}

\newtheorem{theorem}{Theorem}[section]

\newenvironment{proof}{\par\noindent\textit{Proof.}\ }{\hfill$\square$\par}


\definecolor{coauthorEdit}{RGB}{255,0,0}
\definecolor{uniquenessPurple}{RGB}{145,35,180}

\colorlet{blue}{black}
\colorlet{BLUE}{black}
\colorlet{sepTwentyPurple}{black}
\colorlet{newpurple}{black}
\colorlet{NEWPURPLE}{black}
\colorlet{revisiongreen}{black}
\colorlet{revisionorange}{black}
\colorlet{revisionpolish}{black}
\colorlet{REVISIONPOLISH}{black}
\colorlet{coauthorEdit}{black}
\colorlet{uniquenessPurple}{black}
\begin{document}
\title{Generalized Model Fractional Quantum Hall States on Lattices}

\author{Guangyue Ji}
\email{jgy929@pku.edu.cn}
\affiliation{International Center for Quantum Materials, Peking University, Beijing 100871, China}
\affiliation{Beijing Key Laboratory of Quantum Devices, Peking University, Beijing 100871, China}

\author{Jie Wang}
\email{jiewang.phy@pku.edu.cn}
\affiliation{International Center for Quantum Materials, Peking University, Beijing 100871, China}
\affiliation{Beijing Key Laboratory of Quantum Devices, Peking University, Beijing 100871, China}

\begin{abstract}
Model wavefunctions are essential for studying fractional quantum Hall (FQH) phases. They reveal the intricate connections between model FQH states, conformal field theory, Jack polynomials, and pseudopotential Hamiltonians. However, most model wavefunctions studied so far are in the continuum. In this work, we construct model wavefunctions and the corresponding exact parent Hamiltonians for the entire $Z_k$ Read--Rezayi series on a lattice. \textcolor{blue}{Our model wavefunctions obey a generalized clustering rule implemented on the lattice, which we find is encoded in the Macdonald polynomials providing a unified algebraic description of our lattice-specific model states.}
This ensures that the model states are exact zero-energy ground states of local density interactions and exhibit an infinitely large gap in the particle entanglement spectrum. \textcolor{blue}{Numerical studies of the Laughlin and Moore--Read series find the expected zero-mode and one-flux quasihole counting and positive energy gaps at the system sizes studied, and the available finite-size scaling suggests nonzero thermodynamic gaps.} Tuning the model interaction range can drive a phase transition out of the FQH phase. Our work not only advances the theoretical understanding of the underlying mathematical structure of lattice model FQH states but also is potentially useful for designing exotic many-body phases in engineered quantum platforms.
\end{abstract}

\maketitle
\raggedbottom

The fractional quantum Hall (FQH) effect is a paradigmatic example of topological order, whose anyonic excitations provide a promising route toward topological quantum computation. Besides conventional two-dimensional electron gases under strong magnetic fields~\citep{RMP_FQH}, FQH physics and its zero-field relative fractional Chern insulators~\cite{Bernevig_FCI11,LIU2024515,NASS13,Moller_2009} are also actively pursued in emerging systems such as moiré materials and engineered quantum platforms~\citep{Review_FCI_moire26,Clark2020,Léonard2023,Pan_FQH_24}. 
In the study of FQH phases and strongly correlated matter more broadly, model wave functions stand out as an especially powerful tool, inspiring diverse developments ranging from exact parent Hamiltonians to conformal field theory, Jack polynomials, and beyond~\cite{laughlin_anomalous_1983,haldane_fractional_1983,FQH_Review_Conformal17,moore_nonabelions_1991,Bernevig_Jack08,Nayak_2008}.

Model wave functions exhibit ideal properties in several respects, including being exact ground states of certain short-ranged interactions and possessing an infinite gap in the particle entanglement spectrum (PES)~\cite{haldane_fractional_1983,Haldane_EntanglementSpectrum08,Bernevig_PES_11,Pu_LDOS_PES_2022}. These idealized properties have facilitated the design and diagnosis of more general topologically ordered phases in conventional and novel platforms such as moiré flat bands with ideal quantum geometries~\cite{wang_exact_2021,Liu_2025,Ledwith_2023,Mera_KahlerGeo21,OriginModelFCIJie23,HierarchyIdealBandJie22}. These ideal properties are rooted in their distinctive clustering properties, which, generally speaking, are reflected in the asymptotic behavior of the many-body wave function when particles are infinitesimally close to each other~\cite{read_beyond_1999,PairedStateHalfFilling91,simon2020wavefunctionology}. These clustering properties enable Jack polynomials to provide a unified algebraic description of continuum Laughlin and Read--Rezayi model wavefunctions~\cite{Bernevig_Jack08}.

Yet the vast majority of model wavefunctions studied so far are in the continuum. In this work, we systematically construct exact model states for all members of the $Z_k$ Read--Rezayi series on a lattice. \textcolor{blue}{Our lattice states obey a generalized $\delta$-deformed clustering rule, which we show is algebraically encoded in Macdonald polynomials, a generalization of Jack polynomials.} This clustering rule ensures that our states are exact ground states of local density interactions\ with an infinitely large PES gap. Numerical examination of Laughlin and Moore--Read states suggests a finite many-body gap and the expected quasi-hole state counting, further supporting the conclusion that the constructed lattice states are model states. We also investigate phase transitions out of the FQH phase by tuning the interaction range. Given recent progress in engineered flat-band platforms, we expect our theory to provide potentially useful guidance for experimental exploration of lattice FQH states.

\paragraph*{{\bf \em The generalized lattice Laughlin state and parent Hamiltonians.---}}
Our lattice model states are formulated within the Fock space of a lattice lowest Landau level (LLL) with ideal quantum geometry~\cite{wang_exact_2021,Liu_2025,Ledwith_2023,Mera_KahlerGeo21}, as realized, for example, in the Kapit--Mueller model~\citep{Kapit_Exact_2010,shen2025,AtakisiOktel2013,XuPu2020}:
\begin{equation}
    \hat H_{\mathrm{KM}} = |t|\sum_{\bm r,\bm d\in\Lambda} 
    \eta_{\bm d}\, 
    e^{-\frac{|\bm d|^{2}}{4}(\phi^{-1}-1)} 
    e^{\frac{i}{2}\bm d\times\bm r}\, 
    \hat c^{\dagger}_{\bm r-\bm d}\hat c_{\bm r},
    \label{eq:KM}
\end{equation}
where $\hat c^{\dagger}_{\bm r}$ and $\hat c_{\bm r}$ are particle creation and annihilation operators at lattice site $\bm r$. The gauge factor $\eta_{\bm d}=(-1)^{mn+m+n}$ is determined by the lattice vector $\bm d=m\bm a_1+n\bm a_2$. The primitive vectors $\bm a_{1,2}$ enclose flux density $\phi$ per unit cell, and we set the magnetic length to unity. For any $\phi\in(0,1)$, the Kapit--Mueller model has an exactly flat lowest band whose wave functions are lattice samples of holomorphic functions multiplied by the Gaussian factor, $g(z)\exp(-|z|^2/2)$ with $z=(x+iy)/\sqrt2$ for a lattice site at $\bm r=(x,y)$. We set $|t|=1$ throughout.

We start by discussing the lattice Laughlin model state. One of the key results of this section is the following proposed planar $\delta$-deformed Laughlin state at filling $\nu=1/m$,
\begin{equation}
    \Psi_{\{\delta_l\}}^{\nu=\frac{1}{m}}(\{z_i\}) = \prod_{l=1}^{m} \prod_{i<j} (z_i-z_j+\delta_l). \label{eq:laughlin}
\end{equation}

Here $\delta_l$ denotes the complex coordinate of a displacement vector $\bm\delta_l$. The Gaussian factor $\prod_i e^{-\frac12|z_i|^2}$ is implicit. These displacement vectors are chosen in an inversion-symmetric way such that the resulting many-body states respect the inversion symmetry~\cite{haldane_inversion}. If all $\delta_l=0$, Eq.~(\ref{eq:laughlin}) formally reduces to the continuum planar Laughlin wave function, except that the particle coordinates are restricted to lattice sites. The torus generalization is given in Sec.~\ref{sec:supp_laughlin_torus} of the Supplemental Material (SM).

We now explain the motivation for introducing the $\delta$-deformation on a lattice. In the continuum, the Laughlin $\nu=1/m$ model wave function is the exact zero-energy ground state of short-ranged Haldane pseudopotentials~\cite{haldane_fractional_1983,TrugmanKivelson85}. 
The simplest case is the bosonic Laughlin $\nu=1/2$ state, whose parent Hamiltonian is the $v_0$ Haldane pseudopotential; such an interaction can be easily implemented on a lattice as the on-site Hubbard repulsion $U:\hat n_{\bm r}^2:$, with $\hat n_{\bm r}$ being the local density operator. This is why contact-interacting bosons in the Kapit--Mueller model at $\nu=1/2$ have an exact Laughlin ground state~\cite{Kapit_Exact_2010}.

The next simplest case, the fermionic Laughlin $\nu=1/3$ state, already exposes the obstacle. In the continuum, when two fermions approach each other, the many-body wave function vanishes as the third power of their separation, beyond the first-order zero required by Fermi statistics. This clustering feature ensures that the Laughlin $1/3$ state is exactly annihilated by the $v_1$ pseudopotential whose real-space form is given by projecting the Laplacian of the contact interaction $\delta^{(2)}(\bm r)$ into the LLL~\cite{TrugmanKivelson85,haldane_fractional_1983,PairedStateHalfFilling91}. On a lattice, however, fermions never approach infinitesimally close to each other. Therefore the continuum contact-clustering condition has no literal lattice implementation.

\begin{figure}[t]
    \centering
    \includegraphics[width=0.98\columnwidth]{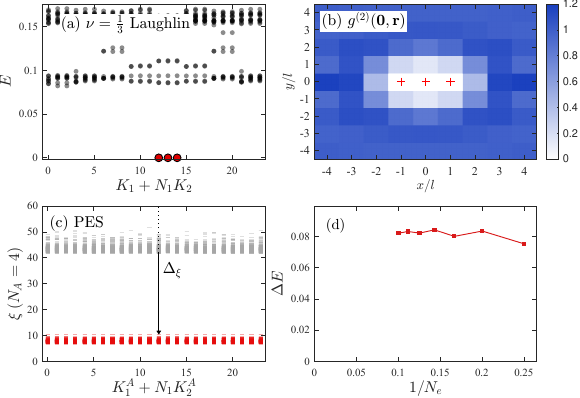}
    \caption{\textbf{Diagnostics of the $\nu=1/3$ $\delta$-deformed lattice Laughlin state.}
    (a) Many-body spectrum of $H_\delta$ for nearest-neighbor interacting fermions at $\nu=\phi=1/3$ in the Kapit--Mueller flat band, computed on a $9\times8$ square lattice. Red dots mark the exactly degenerate threefold zero-energy ground-state manifold.
    (b) Pair correlation function $g^{(2)}(\bm 0,\bm r)$ of one zero mode. The three exact zeros are marked by red crosses.
    (c) Momentum-resolved PES for subsystem size $N_A=4$. An infinitely large entanglement gap separates the universal low-lying levels from nonuniversal levels set by double-precision numerical accuracy~\citep{Behrmann_Model_2016}; the counting below the gap (2730) matches the expected quasihole counting of the Laughlin $\nu=1/3$ state.
    (d) Finite-size scaling of the many-body gap indicating a finite gap in the thermodynamic limit.}
    \label{fig:ED_PES}
\end{figure}

\textcolor{blue}{The displacement deformation in Eqn.~(\ref{eq:laughlin}), which should be regarded as a particular choice of vortex function~\cite{Ledwith_2023}, resolves this problem at the expense of crystal rotational symmetry}~\footnote{\textcolor{blue}{Lattice rotational symmetry can be preserved in some special cases, such as for the $\nu=1/5$ Laughlin state on a square lattice.}}. Instead of requiring an order-$m$ zero at contact, which cannot be directly probed on a lattice, the wave function in Eq.~(\ref{eq:laughlin}) distributes this vanishing condition over prescribed lattice separations, with multiplicities constrained by particle-exchange symmetry. As a direct consequence, this modifies the clustering properties: whenever two particles are separated by one of the prescribed $\bm\delta_l$, the wave function vanishes. Consequently, the $\delta$-deformed wave function is an exact zero-energy state of the density-density interaction~\footnote{
Generally the parent Hamiltonian is a positive linear combination of the corresponding clustering terms $\sum_{\bm r}\sum_l \lambda_l :\hat n_{\bm r}\hat n_{\bm r+\bm\delta_l}:$ with $\lambda_l>0$. This is directly analogous to continuum Haldane pseudopotential Hamiltonians, where arbitrary positive pseudopotential coefficients impose the same excluded relative-angular-momentum channels.}:
\begin{equation}
    \hat H_{\{\bm\delta_l\}} = U \sum_{\bm r}\sum_l :\hat n_{\bm r} \hat n_{\bm r+\bm\delta_l}:, \label{parentH}
\end{equation}
which is derived by matching the interaction pattern to the zero pattern of the many-body wave function. The zero-energy property of the model wave function Eq.~(\ref{eq:laughlin}) under interaction Eq.~(\ref{parentH}) can be numerically confirmed [see Fig.~\ref{fig:ED_PES} (a)]. Despite the fact that the interaction generally breaks lattice rotational symmetry, the prescribed clustering zeros ensure that this interaction exactly annihilates these states. \textcolor{blue}{We emphasize such annihilation does not require band projection, which means the state in Eqn.~(\ref{eq:laughlin}) is an exact ground state for any value of the repulsive interaction strength $U/t$}~\footnote{\textcolor{blue}{The proof is straightforward: first of all, the holomorphic form of these wavefunctions places them in the Kapit-Mueller flat band; the clustering property ensures the expectation value of the interaction is zero. Hence the many-body state is a zero-energy state without requiring band projection.}}. In numerical studies to be presented later, just for simplicity, we project the interaction to the lowest band in the numerical diagnostics below and set $U=1$. For compactness, we refer to the family in Eq.~\eqref{parentH} as $H_\delta$ below.

In closing this section, we comment that the continuum form of the trial wave function in Eq.~(\ref{eq:laughlin}) has been proposed as a variational ansatz for studying symmetry-breaking states in conventional Landau levels~\cite{Fogler04,MusaelianJoynt96,WexlerCiftja02,CiftjaLapilliWexler04,Qiu_2012,Wang_Wan_Zhang_2012,Pu_2024}. However, it was not previously recognized that such a wave function constitutes a model state on a lattice. We further point out the $\delta$-deformed Laughlin wave function and its non-Abelian counterparts can be systematically constructed on the plane using Macdonald polynomials, a generalization of Jack polynomials. We develop this construction in the last section of this work.

\paragraph*{{\bf \em Spectrum, entanglement, and correlations of the generalized lattice Laughlin state.---}} We now examine other characteristics of the lattice Laughlin state, including its real-space correlations, PES, and finite-size scaling of the many-body gap. In this section, we focus on the fermionic $\nu=1/3$ state as the representative example. Numerical studies of other topologically ordered states, including the $\nu=1/4$ Laughlin state and non-Abelian states, are presented in the following sections and in the SM~\cite{supplemental}.

We consider nearest-neighbor interacting fermions at $\nu=1/3$ in the Kapit--Mueller flat band, with interaction $\sum_{\bm r} \hat n_{\bm r}\hat n_{\bm r\pm\bm a_1}$. This corresponds to setting $\{ \bm \delta_l\} = \{0,\pm\bm a_1\}$ in Eq.~(\ref{parentH}). The many-body spectrum exhibits an exactly threefold degenerate zero-energy manifold separated by a finite gap, as shown in Fig.~\ref{fig:ED_PES}(a). These three zero modes permute among themselves under twisted boundary flux insertion, as demonstrated in Sec.~\ref{sec:supp_twist_flow_subsection} of the SM.

The pair-correlation function $g^{(2)}$ directly visualizes the clustering property. 
Figure~\ref{fig:ED_PES}(b) shows
$g^{(2)}(\bm r_0,\bm r) = \langle :\hat{n}_{\bm r_0} \hat{n}_{\bm r}: \rangle /[ \langle \hat{ n}_{\bm r_0}\rangle \langle \hat{ n}_{\bm r}\rangle ]$,
computed from one of the three zero modes. The exact zeros at the origin and $\pm\bm a_1$ reveal the imposed lattice clustering condition. Since the many-body energy is the two-body interaction weighted by this pair correlation, the nodal pattern directly explains the zero-energy nature of the model states.

We further compute the PES. Fig.~\ref{fig:ED_PES}(c) shows the momentum-resolved PES for the lattice model $\nu=1/3$ state, from which an infinitely large entanglement gap is clearly visible. The counting below this gap matches that of quasihole excitations, again supporting that this state is a model state~\cite{Bernevig_PES_11}. Finally, Fig.~\ref{fig:ED_PES}(d) shows the finite-size scaling of the many-body gap, suggesting a finite many-body gap in the thermodynamic limit.

\paragraph*{{\bf \em Interaction-range-induced transition.---}}
In this section, we show that when increasing the interaction range of $H_\delta$, phase transitions can occur, even though the corresponding model states remain exact zero-energy ground states by construction. The transition is signaled by the closing of the many-body spectral gap, the emergence of a finite entanglement gap, and the loss of the Laughlin quasihole counting structure in the low-lying entanglement spectrum.

\begin{figure}[t]
\centering
\includegraphics[width=0.98\columnwidth]{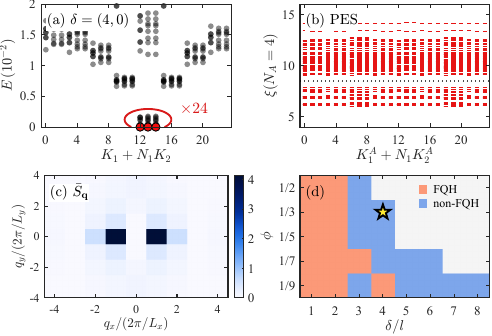}
\caption{
\textbf{Interaction-range-induced phase transitions.}
(a) Many-body spectrum at $\nu=\phi=1/3$ for $N_e=8$ with interaction displacement $\bm\delta=4\bm a_1$. Additional low-energy states, highlighted by the red ellipse, appear at large $|\bm\delta|$, indicating the closing of the Laughlin many-body gap.
(b) Corresponding PES for $N_A=4$. A finite principal entanglement gap opens, marked by the dotted line, which sets the cutoff used to identify the low-lying PES manifold.
(c) Projected structure factor $\bar S(\bm q)$ of a representative zero-energy state, showing pronounced finite-momentum peaks.
(d) Phase diagram obtained from the combined diagnostics of the energy spectrum, PES counting, and projected structure factor. Here $l=|\bm a_1|=|\bm a_2|$ denotes the lattice spacing. The star marks the representative point shown in panels (a)--(c), and the underlying numerical data are summarized in Sec.~\ref{sec:supp_phase_diagram_ne8} of SM.
The calculations are performed for $N_e=8$ on rectangular tori with aspect ratios chosen as close to unity as allowed by flux commensurability.
} \label{fig:phase_diagram}
\end{figure}

As a case study, we examine fermions at $\nu=1/3$ in the Kapit--Mueller model with $\phi=1/3$. The interaction uses the displacement $\bm\delta=\delta\bm a_1$, i.e., $\hat n_{\bm r}\hat n_{\bm r\pm\delta\bm a_1}$, and we track the many-body spectrum, PES, and projected structure factor as the scalar range $\delta$, measured in lattice spacings, is increased. Representative results for the $N_e=8$ system are shown in Fig.~\ref{fig:phase_diagram}.

As the interaction range increases, the low-energy manifold changes qualitatively. 
For $\bm\delta=4\bm a_1$, additional low-energy states appear beyond the three exact zero modes [Fig.~\ref{fig:phase_diagram}(a)], with the near-degenerate manifold reaching 24 states, equal to the number of magnetic unit cells. 
Although the three zero modes of the corresponding $H_\delta$ can still be described within the generalized wave-function framework, the PES develops finite principal gaps [Fig.~\ref{fig:phase_diagram}(b)] and the counting of low-lying spectral modes deviates from the expected Laughlin quasihole counting. At the same time, the projected structure factor $\bar S(\bm q) = \frac{1}{N_e} \left[ \langle \hat{\bar n}_{-\bm q}\hat{\bar n}_{\bm q}\rangle - \langle \hat{\bar n}_{-\bm q}\rangle \langle \hat{\bar n}_{\bm q}\rangle \right]$ defined in terms of the flatband projected density operator $\hat{\bar n}$ develops pronounced finite-momentum peaks [Fig.~\ref{fig:phase_diagram}(c)], indicating the onset of density-modulated correlations and translation symmetry breaking.

\paragraph*{{\bf \em Non-Abelian generalizations.---}}
We now extend our analytical and numerical discussions to non-Abelian states in the Read--Rezayi $Z_k$ series~\cite{read_beyond_1999}. We organize this section as follows. We first discuss the general construction of our lattice-model wavefunctions, including both Abelian and non-Abelian cases derived from Macdonald polynomials. We then present a numerical exploration of the Moore-Read and Fibonacci states. 

Macdonald polynomials are a generalization of Jack polynomials, and it is known that Jack polynomials provide a unified algebraic description for many continuum FQH model wavefunctions~\cite{Bernevig_Jack08}. In what follows, we start with a brief review of Jack polynomials and continuum model FQH wavefunctions.
Jack polynomials form a basis for symmetric polynomials at generic parameter values and, for suitable parameter values and root configurations, satisfy prescribed clustering conditions. The model wavefunctions of the bosonic Read--Rezayi series at filling $\nu=k/2$ are generated from the densest root occupation $k0k0\cdots k$ for \mbox{$N=kM$}, satisfying the $(k,2)$ admissibility rule: at most $k$ particles in any two consecutive orbitals. The associated Jack polynomial is obtained from this root configuration by solving the Laplace--Beltrami eigenvalue problem at $\alpha=-(k+1)$. Multiplying it by the $p$th power of the Vandermonde determinant gives a state at filling $\nu=k/(kp+2)$.

Our construction of lattice model states follows the same spirit as the continuum construction, but with modified clustering properties. First of all, we introduce a symmetric $N$-variable function $\mathcal W_{N,\delta}^{(k)}(\bm z)$ at filling $k/2$ dictated by the $(k+1)$-body $\delta$-deformed lattice clustering rule. Later, we explain how this function is deduced from Macdonald polynomials. Explicitly, $\mathcal W_{N,\delta}^{(k)}(\bm z)$ satisfies
{\color{blue}
\begin{equation}
    \mathcal W_{N,\delta}^{(k)}(\ldots,z_0,z_0+\delta,\ldots,z_0+k\delta,\ldots)=0,
    \label{eq:wheel}
\end{equation}
}
for arbitrary positions of the remaining particles. The function $\mathcal W$ naturally describes bosonic lattice model states. With it in hand, we show \textcolor{blue}{the entire wavefunction at $\nu=k/(kp+2)$ is obtained by multiplying $\mathcal W$ by a displaced Jastrow factor, yielding a fermionic (bosonic) state for odd (even) $p$,}
{\color{blue}
\begin{equation}
\begin{aligned}
    \Psi_{N,\delta,\delta'}^{(k,p)}(\bm z) &= \mathcal J_{N,\delta'}^{(p)}(\bm z)\mathcal W_{N,\delta}^{(k)}(\bm z),\\
    \mathcal J_{N,\delta'}^{(p)}(\bm z) &= \prod_{a=1}^{p}\prod_{i<j}(z_i-z_j+\delta'_a),
\end{aligned} \label{eq:jrr_dressed}
\end{equation}
}
where the displacement set $\{\delta'_a\}$ is chosen to be invariant under inversion to ensure statistics.

For example, when $k=p=1$ and $\delta'=0$, the state $\Psi_{N,\delta,0}^{(1,1)}(\bm z)$ is precisely the $\delta-$deformed Laughlin $1/3$ wavefunction in Eq.~(\ref{eq:laughlin}) discussed at the beginning. This is immediately seen by noticing that $\mathcal W_{N,\delta}^{(1)}(\bm z)\propto\prod_{i<j}[(z_i-z_j)^2-\delta^2]$. 

Last but not least, we show how $\mathcal W_{N,\delta}^{(k)}$ is generated from Macdonald polynomials. Macdonald polynomials are a two-parameter $(t_{\rm M}, q_{\rm M})$ generalization of Jack polynomials. \textcolor{blue}{For $(k,2)$-admissible root configurations, Jack polynomials at $\alpha=-(k+1)$ satisfy $(k+1)$-particle contact-clustering conditions; instead, the Macdonald polynomials at $q_{\rm M}=t_{\rm M}^{-(k+1)}$ satisfy the \emph{wheel condition}: they vanish whenever any $k+1$ variables take the values $x,t_{\rm M}x,\ldots,t_{\rm M}^{k}x$, for arbitrary values of the remaining variables~\cite{FeiginMacdonaldWheel2003}. For the densest root $\lambda_k^{(N)}$ corresponding to $k0k0\cdots k$, we denote the corresponding Macdonald polynomial by $\widehat P_{\lambda_k^{(N)}}(\bm x;t_{\rm M})$.}

\textcolor{blue}{An exponential change of variables maps the wheel condition to the $\delta$-deformed lattice clustering rule. Set $x_i=e^{\epsilon\xi_i}$ and $t_{\rm M}=e^{\epsilon\delta}$, where $\xi_i=z_i-N^{-1}\sum_jz_j$ is the coordinate measured from the center of mass (COM). Since $z_j-z_i=\delta$ implies $x_j=t_{\rm M}x_i$, any $k+1$ particles successively separated by $\delta$ satisfy the wheel condition for every $\epsilon$. Each coefficient in the Taylor expansion in $\epsilon$ therefore satisfies the same clustering condition. With $\delta$ held fixed, the first nonzero coefficient defines the clustering polynomial of lowest total degree:
{\begin{equation}
    \widehat P_{\lambda_k^{(N)}}\! \left(e^{\epsilon\bm\xi};e^{\epsilon\delta}\right) = \frac{\epsilon^{D_{N,k}}}{\mathcal N_{N,k}} \mathcal W_{N,\delta}^{(k)}(\bm z) + O\!\left(\epsilon^{D_{N,k}+1}\right),\label{eq:macdonald_leading}
\end{equation}}}
where $e^{\epsilon\bm\xi}$ denotes $(e^{\epsilon\xi_1},\ldots,e^{\epsilon\xi_N})$, $D_{N,k}=N(N-k)/k$ is the total degree in $(\bm z,\delta)$, and $\mathcal N_{N,k}$ is the normalization. This explains how Macdonald polynomials generate the $\delta$-deformed clustering functions $\mathcal W$. The deformation from continuum contact clustering to discrete lattice clustering is encoded in the Macdonald wheel condition expressed in exponential coordinates.

\textcolor{blue}{As proved in Sec.~\ref{sec:supp_minimal_degree} of the SM, within the stated symmetry class, $\mathcal W_{N,\delta}^{(k)}$ uniquely satisfies the clustering rule at the same minimum degree as the contact Read--Rezayi Jack, up to normalization.}

\begin{figure}[H]
    \centering
    \begin{overpic}[width=\columnwidth]{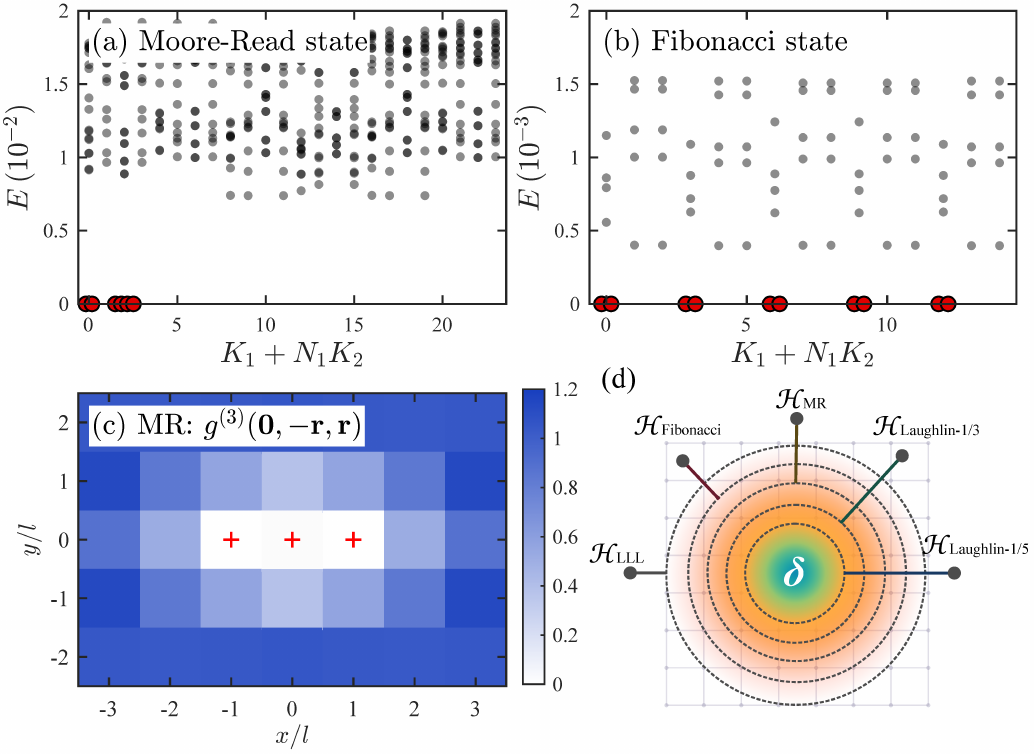}
    \setlength{\fboxsep}{0.5pt}
    \put(12,49){\colorbox{white}{\tiny $N=12,\ N_\phi=24,\ 8\bm a_1\!\times\!6\bm a_2$}}
    \put(63,68){\colorbox{white}{\fontsize{7}{8}\selectfont Fibonacci \textcolor{blue}{candidate}}}
    \put(61,49){\colorbox{white}{\tiny $N=9,\ N_\phi=15,\ 6\bm a_1\!\times\!5\bm a_2$}}
    \end{overpic}
    \caption{
    \textbf{Diagnostics of lattice Moore--Read and Fibonacci states.}
    (a) Many-body spectrum of the fermionic Moore--Read state at $\nu=1/2$, generated by the three-body interaction
    $\hat n_{\bm r} \hat n_{\bm r+\bm a_1}\hat n_{\bm r+2\bm a_1}$.
    (b) Many-body spectrum of the fermionic Fibonacci \textcolor{blue}{candidate} at $\nu=3/5$, generated by the four-body interaction
    $\hat n_{\bm r} \hat n_{\bm r+\bm a_1}\hat n_{\bm r+2\bm a_1}\hat n_{\bm r+3\bm a_1}$.
    The zero-energy manifolds, highlighted in red, contain six and ten states, respectively.
    (c) Three-particle correlation function
    $g^{(3)}(\bm 0,-\bm r,\bm r)$ for the Moore--Read state.
    The suppressed weight along $\bm r=(\pm1,0)$ reflects the imposed three-body clustering constraint; red crosses mark the corresponding correlation zeros.
    (d) Schematic hierarchy of the null spaces of model Hamiltonians obtained by increasing the interaction cluster size or by reducing the number of interaction constraints~\cite{Yuzhu2023}.
    }
    \label{fig:lattice_fMR}
\end{figure}

\textcolor{blue}{By construction, the wavefunction in Eq.~\eqref{eq:jrr_dressed} is an exact zero-energy ground state of the positive-semidefinite parent Hamiltonian,}
{\color{blue}
\begin{equation}
\widehat H_{\rm RR}=\sum_{\bm R}\sum_{a=1}^{p}:\!\hat n_{\bm R}\hat n_{\bm R+\bm\delta'_a}\!:+\sum_{\bm R}:\!\hat n_{\bm R}\hat n_{\bm R+\bm\delta}\cdots\hat n_{\bm R+k\bm\delta}\!:, \label{HdeltaRR}
\end{equation}
}
where the two terms enforce the two-particle and $(k+1)$-particle clustering conditions, respectively.  The fermionic Moore--Read and Fibonacci systems at $\nu=1/2$ and $3/5$, respectively, are examined numerically in Fig.~\ref{fig:lattice_fMR}(a),(b). \textcolor{blue}{Finite-size scaling further suggests a nonzero thermodynamic energy gap for Moore--Read at $\nu=1/2$ (see Fig.~\ref{fig:supp_mr_gap_scaling} of the SM~\cite{supplemental}). The Fibonacci state shows stronger finite-size effects, leaving its thermodynamic gap unresolved; for this reason, we call it a ``Fibonacci candidate''.}

Further numerical support comes from zero-mode and one-flux quasihole counting. For the systems shown in Fig.~\ref{fig:lattice_fMR}(a),(b), we find six and ten exact zero modes, respectively, matching the expected finite-torus counts for the Moore--Read and Fibonacci states at these sizes. \textcolor{blue}{Further finite-size zero-mode and one-flux quasihole counts are given in Secs.~\ref{sec:supp_laughlin_numerics}, \ref{sec:supp_general_k_kernel}, and \ref{sec:supp_fibonacci_torus} of the SM~\cite{supplemental}; the Fibonacci counts also match continuum predictions in several cases, though some small systems have extra zero modes.}

\paragraph*{{\bf \em Summary and outlook.---}}
In this work, we systematically generalized continuum FQH model wave functions to lattices for all Read--Rezayi $Z_k$ families.  Our lattice-model wavefunctions obey a lattice-specific clustering rule encoded in Macdonald polynomials, which generalize Jack polynomials. By construction, these wavefunctions are exact zero-energy states of local density interactions. Numerical calculations support the conclusion that they are gapped, topologically ordered states when the deformation parameter is below the critical value, most conclusively for the Laughlin series and the Moore-Read state at $\nu=1/2$.

Finally, we highlight a few novel features of our lattice-model states and provide an outlook on future work. First, the construction of our model states does not require band projection, which means that they are exact ground states even for infinitely large interactions. \textcolor{blue}{Bosons with an infinite on-site interaction are naturally represented by spins; hence, we envision that this insight could be used to find model Hamiltonians for chiral spin liquids such as the Kalmeyer-Laughlin spin liquid~\cite{KalmeyerLaughlin1987,Kalmeyer1989,Kapit_Exact_2010,ShenJiWangInPreparation}.} 
\textcolor{blue}{Moreover, the algebraic description of our model state based on Macdonald polynomials may enable systematic studies of the nature of collective modes and anyonic excitations in lattice FCIs~\cite{MannaWildeboerSierraNielsen2018,MannaWildeboerNielsen2019}, following the example of Jack polynomials, which have been known to provide key insights into them in the continuum limit.}  Extending our theory to other topological orders such as unprojected Jain states~\cite{RezayiMacDonald1991Jain,BandyopadhyayJainParent2020} constitutes one important future direction. \textcolor{blue}{The displaced Jain state is an exact zero mode of the lattice density interaction, but finite-torus tests find additional zero modes whose origin deserves further study; see Sec.~\ref{sec:supp_jain_multiband} of the SM for details.} We also believe the $\delta-$tuned phase transition between topological order and trivial phases deserves further study~\cite{26_Sodemann_Plasma,26_Lu_Ziyang_Transition,26_Song_PhaseTransition}. Lastly, we expect that our model states will be beneficial for searching for fractionalized states on engineered platforms~\cite{Léonard2023,Clark2020,Pan_FQH_24}. Promising near-future targets include the fermionic Laughlin $\nu=1/3$ state, the bosonic Laughlin $\nu=1/4$ state, and non-Abelian lattice FQH states. 

\let\SMsavedaddcontentsline\addcontentsline
\renewcommand{\addcontentsline}[3]{}
\section*{Acknowledgments}
We acknowledge Jainendra K. Jain, Zhao Liu, Songyang Pu, Yuzhu Wang, Jingxin Liu, Guangyu Yu, Bo Yang, Junren Shi and Xincheng Xie for useful discussions. We acknowledge Jinjie Zhang and Xin Shen for ongoing collaborations on related topics.

We acknowledge support from Quantum Science and Technology-National Science and Technology Major Project (Grant No. 2025ZD0300500) and The Fundamental Research Funds for the Central Universities, Peking University.

\expandafter\def\csname OriginalNewBib@BandyopadhyayJainParent2020\endcsname{1}
\expandafter\def\csname OriginalNewBib@FeiginJackIdeal2002\endcsname{1}
\expandafter\def\csname OriginalNewBib@FeiginMacdonaldWheel2003\endcsname{1}
\expandafter\def\csname OriginalNewBib@JolicoeurLuque2011HighestWeight\endcsname{1}
\expandafter\def\csname OriginalNewBib@MatsubaraMatsuda1956\endcsname{1}
\expandafter\def\csname OriginalNewBib@RezayiMacDonald1991Jain\endcsname{1}
\expandafter\def\csname OriginalNewBib@ShenJiWangInPreparation\endcsname{1}

\let\reviewbibitem\bibitem
\def\reviewbibsetup{%
 \let\reviewbibinfo\bibinfo
 \renewcommand{\bibinfo}[2]{%
  \def\reviewfield{##1}\def\reviewauthor{author}\def\reviewnote{note}%
  \def\reviewauthorskey{Yuzhu2023}\def\reviewnotekey{supplemental}%
  \ifx\reviewkey\reviewauthorskey
   \ifx\reviewfield\reviewauthor
    \reviewbibinfo{##1}{##2}%
   \else\reviewbibinfo{##1}{##2}\fi
  \else\ifx\reviewkey\reviewnotekey
   \ifx\reviewfield\reviewnote
    \reviewbibinfo{##1}{See Supplemental Material for additional results for lattice model states; interaction-range-induced phase transitions; the Macdonald-polynomial construction; and the extension to unprojected Jain states.}%
   \else\reviewbibinfo{##1}{##2}\fi
  \else\reviewbibinfo{##1}{##2}\fi\fi}%
 \let\reviewbibsetup\relax}
\renewcommand{\bibitem}[2][]{%
 \reviewbibsetup
 \def\reviewkey{#2}%
 \reviewbibitem[#1]{#2}}

\let\bibitem\reviewbibitem
\let\bibinfo\reviewbibinfo
\let\addcontentsline\SMsavedaddcontentsline

\clearpage
\onecolumngrid
\begin{center}
\textbf{Supplemental Material for ``Generalized Model Fractional Quantum Hall States on Lattices''}
\end{center}

\setcounter{equation}{0}
\setcounter{figure}{0}
\setcounter{table}{0}
\setcounter{page}{1}
\setcounter{section}{0}
\setcounter{secnumdepth}{2}

\makeatletter
\renewcommand{\theequation}{S\arabic{equation}}
\renewcommand{\thefigure}{S\arabic{figure}}
\renewcommand{\thesection}{S\Roman{section}}
\renewcommand{\thepage}{\arabic{page}}
\renewcommand{\thetable}{S\arabic{table}}
\renewcommand{\theHequation}{S\arabic{equation}}
\renewcommand{\theHfigure}{S\arabic{figure}}
\renewcommand{\theHsection}{S\arabic{section}}
\renewcommand{\theHtable}{S\arabic{table}}
\makeatother

\tableofcontents

This Supplemental Material is organized into four sections. Section~\ref{sec:supp_additional_numerics} collects additional results for the Laughlin and non-Abelian states, including torus wavefunctions, numerical diagnostics, interaction symmetry, the small-flux continuum limit, and hopping-truncation benchmarks. Section~\ref{sec:supp_phase_diagram_ne8} presents the interaction-range-induced phase transitions. Section~\ref{sec:supp_recursive_mr} provides a more detailed discussion on the Macdonald-polynomial construction and its proofs. Section~\ref{sec:supp_jain_multiband} explores the extension to unprojected Jain states.

\section{Additional results for lattice model states}
\label{sec:supp_additional_numerics}
\label{sec:supp_twist_flow}

\subsection{Torus Laughlin wavefunctions}
\label{sec:supp_laughlin_torus}

The torus generalization of Eq.~\eqref{eq:laughlin} uses the modified Weierstrass sigma function $\sigma(z)$, defined on the torus~\cite{haldanetorus1,haldanetorus2,haldane2018origin,Wang_2019}. For $N$ particles and $N_\phi=mN$ flux quanta, we define $f(z)=\sigma(z)e^{-|z|^2/(2N_\phi)}$ using the complex-coordinate convention of the main text. The wavefunction takes the form
\begin{equation}
 \Psi_{\{\delta_l\},\alpha}^{\nu=1/m}(\{z_i\})
 =\prod_{l=1}^{m}\prod_{i<j}f(z_i-z_j+\delta_l)
  \prod_{a=1}^{m}f(Z-\alpha_a),
 \qquad Z=\sum_i z_i .
 \label{eq:supp_laughlin_torus}
\end{equation}
Here $\alpha_a$ are the center-of-mass zeros. Following the periodic boundary conditions used in Ref.~\cite{Wang_2019}, we choose representatives satisfying $\sum_a\alpha_a=0$. For fixed magnetic boundary conditions, the allowed center-of-mass factors form an $m$-dimensional space~\cite{NiuWenGSD,haldanetorus1,haldanetorus2}. The inversion-symmetric displacement set imposes the prescribed pair zeros. Setting all displacements to zero recovers the usual torus Laughlin state, while evaluating the wavefunction at lattice sites gives its lattice counterpart.

For a lattice torus, $L_x$ and $L_y$ count sites along $\bm a_1$ and $\bm a_2$, respectively. Boundary angles $(\theta_x,\theta_y)$ are the total phases along the two physical torus cycles after removing the magnetic gauge phase, modulo $2\pi$. Unless otherwise specified, when $q$ divides $L_x$, the numerical calculations use the reference angles $(\theta_x^{\rm ref},\theta_y^{\rm ref})=\pi[(L_x/q)\bmod2,L_y\bmod2]$ at $\phi=1/q$. If $q$ divides $L_y$ but not $L_x$, the magnetic cell is taken along $y$, giving $\pi[L_x\bmod2,(L_y/q)\bmod2]$ in the same physical coordinates; interaction directions remain fixed. Additional flux insertion is denoted by $\Delta\bm\theta=\bm\theta-\bm\theta^{\rm ref}$. Other boundary choices are stated with the corresponding results.

\subsection{\texorpdfstring{Spectral flow of the Laughlin $\nu=1/3$ ground-state manifold}{Spectral flow of the Laughlin nu=1/3 ground-state manifold}}
\label{sec:supp_twist_flow_subsection}

In this subsection, we show that the three zero modes of the fermionic $\nu=1/3$ lattice Laughlin manifold permute among themselves under twisted boundary flux insertion. We add a small transverse interaction to split the three exact zero modes and then track the resulting low-energy states as a function of the inserted flux. The resulting smooth three-state spectral flow is consistent with Laughlin topological order.

\begin{figure}[htbp]
    \centering
    \includegraphics[width=0.5\linewidth]{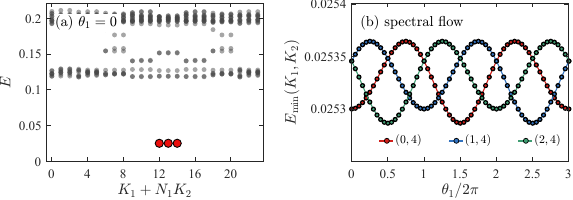}
    \caption{
    \textbf{Spectral flow under twisted boundary conditions.}
    (a) Many-body spectrum at $\theta_1=0$ for $N_e=8$ fermions at $\nu=\phi=1/3$ on a $9\bm a_1\times 8\bm a_2$ lattice. The interaction is
    $2(\sum_{\bm r}\hat n_{\bm r}\hat n_{\bm r+\bm a_1}
    +0.1\sum_{\bm r}\hat n_{\bm r}\hat n_{\bm r+\bm a_2})$,
    projected to the lattice lowest Landau level. Red dots mark the three lowest states.
    (b) Spectral flow of the lowest-energy states in the three momentum sectors $(K_1,K_2)=(0,4),(1,4),(2,4)$, under the additional twist $\theta_1=\Delta\theta_x$, with $\Delta\theta_y=0$. The smooth evolution and permutation of these three low-energy states under flux insertion support their Laughlin-type topological character, even though the weak transverse interaction removes the exact zero-energy condition.
    }
    \label{fig:supp_twist_flow}
\end{figure}

We implement this perturbation by adding a weak density-density interaction in the transverse lattice direction. The projected interaction is
\begin{equation}
    H_{\mathrm{Per}}
    =
    2\mathcal P_{\mathrm{LLL}}
    \left[
    \sum_{\bm r} \hat n_{\bm r}\hat n_{\bm r+\bm a_1}
    +0.1\sum_{\bm r} \hat n_{\bm r}\hat n_{\bm r+\bm a_2}
    \right]
    \mathcal P_{\mathrm{LLL}} .
\end{equation}
Unlike the exact parent Hamiltonian with only the $\bm a_1$ interaction, this perturbation lifts the exact zero-energy condition. Nevertheless, the three lowest states remain well separated from higher excited states. As shown in Fig.~\ref{fig:supp_twist_flow}, the three lowest states evolve smoothly and permute under flux insertion, consistent with the expected topological spectral flow of the Laughlin phase.

\subsection{Laughlin states on the torus}
\label{sec:supp_laughlin_numerics}
\label{sec:supp_boson_laughlin14}
We compare the fermionic $\nu=1/3$ and bosonic $\nu=1/4$ lattice Laughlin parents at flux density $\phi=1/3$. In both cases the interaction is projected into the Kapit--Mueller lowest band for the numerical diagnostics. The fermionic parent is $H^{\rm f}_{1/3}=U\sum_{\bm r}\hat n_{\bm r}(\hat n_{\bm r+\bm a_1}+\hat n_{\bm r-\bm a_1})$, as in the main text. Its three zero modes, pair correlations, and PES are shown in Fig.~\ref{fig:ED_PES}; the spectral flow is discussed in Sec.~\ref{sec:supp_twist_flow_subsection}. The bosonic parent additionally imposes the on-site exclusion. With $U=1$, it is
\begin{equation}
H^{\rm b}_{1/4}
=
\sum_{\bm r} :\hat n_{\bm r}\hat n_{\bm r}:
+
\sum_{\bm r}
\left(
\hat n_{\bm r}\hat n_{\bm r+\bm a_1}
+
\hat n_{\bm r}\hat n_{\bm r-\bm a_1}
\right).
\label{eq:supp_boson14_H}
\end{equation}
The first term is the normal-ordered on-site Hubbard repulsion. 
The second term imposes nearest-neighbor exclusions along the $\bm a_1$ direction. 
Both $\pm\bm a_1$ bonds are written explicitly to match the convention used in the numerical implementation; this convention fixes the energy normalization but does not affect the zero-mode condition.

\begin{figure}[htbp]
    \centering
    \includegraphics[width=1\textwidth]{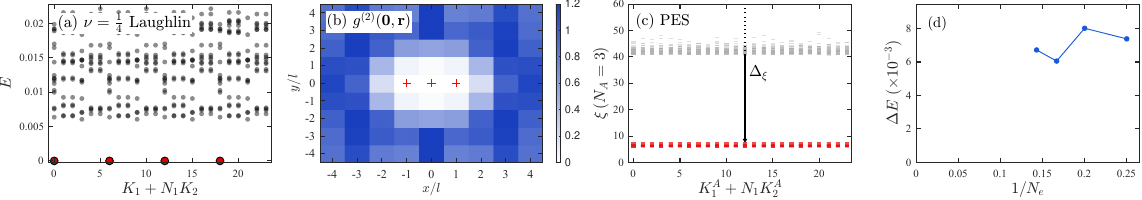}
    \caption{
    \textbf{Diagnostics of the bosonic $\nu=1/4$ $\delta$-deformed lattice Laughlin state.}
    (a) Many-body spectrum of Eq.~(\ref{eq:supp_boson14_H}) for $N_e=6$ bosons in the Kapit--Mueller flat band with $\phi=1/3$, computed on a $9\bm a_1\times 8\bm a_2$ lattice. 
    Red dots mark the exactly degenerate fourfold zero-energy ground-state manifold.
    (b) Pair-correlation function $g^{(2)}(\bm 0,\bm r)$ of a representative zero mode. 
    Red crosses mark the exact correlation zeros.
    (c) Momentum-resolved PES for subsystem size $N_A=3$. 
    The PES is consistent with an infinite entanglement gap within double-precision numerical accuracy; the counting below the gap matches the expected quasihole counting (728).
    (d) Finite-size scaling of the many-body gap.
    }
    \label{fig:supp_boson_laughlin14}
\end{figure}

The diagnostics in Fig.~\ref{fig:supp_boson_laughlin14} confirm that this state is the bosonic counterpart of the $\nu=1/3$ lattice Laughlin state discussed in the main text. 
The spectrum contains an exactly fourfold degenerate zero-energy manifold separated from excited states by a finite gap. 
The pair-correlation function vanishes at the origin and at the two nearest-neighbor positions selected by the interaction geometry. The on-site repulsion requires a contact zero, which is at least quadratic because bosonic exchange symmetry permits only even powers of the relative coordinate. These three positions correspond to four wavefunction zeros counting multiplicity: a double zero at the origin and a simple zero at each displaced position. The PES displays an infinitely large entanglement gap up to double-precision numerical accuracy, and the number of levels below the gap agrees with the quasihole counting of the bosonic Laughlin $\nu=1/4$ state. 
The many-body gap remains finite for the accessible system sizes, as shown by the finite-size scaling in Fig.~\ref{fig:supp_boson_laughlin14}(d).

\emph{Finite-size zero modes and gaps.---}\textcolor{blue}{The tested fermionic ($N=4,\ldots,10$) and bosonic ($N=4,\ldots,7$) Laughlin systems have three and four zero modes, respectively, matching the continuum torus counts (Table~\ref{tab:supp_laughlin_counts}). All listed systems have a positive energy gap above the zero-mode manifold.}

\begin{table}[H]
\caption{Laughlin zero-mode counts and gaps at the parent fillings $\nu=1/m$ and flux density $\phi=1/3$. Throughout the following tables, $L_x\times L_y$ gives the physical site counts, $D_{\rm lat}$ and $D_{\rm cont}$ denote the lattice and continuum zero-mode counts summed over all momentum sectors, and $\Delta_+$ is the energy gap above the full zero-mode manifold. The interactions and their normalization are defined in this subsection.}
\label{tab:supp_laughlin_counts}
\centering\small
\begingroup
\begin{ruledtabular}
\begin{tabular}{ccccccc}
$\nu$ & $N$ & $N_\phi$ & $L_x\times L_y$ & $D_{\rm lat}$ & $D_{\rm cont}$ & $\Delta_+/U$\\\hline
\multicolumn{7}{l}{\textbf{Fermions, $\nu=1/3$}}\\[2pt]
$1/3$ & 4 & 12 & $6\times 6$ & 3 & 3 & $7.54281\times10^{-2}$\\
$1/3$ & 5 & 15 & $9\times 5$ & 3 & 3 & $8.36389\times10^{-2}$\\
$1/3$ & 6 & 18 & $9\times 6$ & 3 & 3 & $8.02379\times10^{-2}$\\
$1/3$ & 7 & 21 & $9\times 7$ & 3 & 3 & $8.43983\times10^{-2}$\\
$1/3$ & 8 & 24 & $9\times 8$ & 3 & 3 & $8.23232\times10^{-2}$\\
$1/3$ & 9 & 27 & $9\times 9$ & 3 & 3 & $8.33618\times10^{-2}$\\
$1/3$ & 10 & 30 & $9\times 10$ & 3 & 3 & $8.22551\times10^{-2}$\\
\hline
\multicolumn{7}{l}{\textbf{Bosons, $\nu=1/4$}}\\[2pt]
$1/4$ & 4 & 16 & $6\times 8$ & 4 & 4 & $7.38296\times10^{-3}$\\
$1/4$ & 5 & 20 & $6\times 10$ & 4 & 4 & $8.01592\times10^{-3}$\\
$1/4$ & 6 & 24 & $9\times 8$ & 4 & 4 & $6.05529\times10^{-3}$\\
$1/4$ & 7 & 28 & $12\times 7$ & 4 & 4 & $6.71932\times10^{-3}$\\
\end{tabular}
\end{ruledtabular}
\endgroup
\end{table}

\emph{Continuum quasihole reference.---}At $N_\phi=mN+s$, the continuum Laughlin zero modes are counted by periodic orbital configurations with at most one particle in any $m$ consecutive orbitals.\footnote{B. A. Bernevig and N. Regnault, \href{https://doi.org/10.1103/PhysRevB.85.075128}{Phys. Rev. B \textbf{85}, 075128 (2012)}.}
Writing each cyclic spacing between occupied orbitals as $m+t_i$, with $t_i\ge0$ and $\sum_i t_i=s$, gives $\binom{N+s-1}{s}$ choices. Counting a marked particle at any of the $N_\phi$ orbitals and dividing by the $N$ choices of marked particle yields
{
\begin{equation}
D_{\rm cont}^{\rm Laughlin}(N,N_\phi)
=\frac{N_\phi}{N}\binom{N+s-1}{s},
\qquad N_\phi=mN+s,\quad s\ge0.
\label{eq:supp_laughlin_qh_count}
\end{equation}
}
This count includes all translations and momentum sectors. It gives $D_{\rm cont}=m$ at the parent filling and $D_{\rm cont}=N_\phi$ after adding one flux quantum. Applied to the particle cut, with $N$ replaced by $N_A$ at fixed $N_\phi$, it gives $2730$ for $(m,N_A,N_\phi)=(3,4,24)$ and $728$ for $(4,3,24)$, matching the PES counts shown in Figs.~\ref{fig:ED_PES}(c) and \ref{fig:supp_boson_laughlin14}(c).

\emph{Direct lattice quasihole counting.---}We add one flux quantum at fixed particle number, $N_\phi=mN+1$, and diagonalize the same lattice parent interactions in all momentum sectors. The physical boundary twists are kept equal to those of the corresponding parent systems in Table~\ref{tab:supp_laughlin_counts}. The quasihole geometries are chosen with $L_x\ge L_y$ and aspect ratio as close to unity as the integer site counts allow. Table~\ref{tab:supp_laughlin_qh} compares the lattice zero-mode counts with Eq.~\eqref{eq:supp_laughlin_qh_count}.

\textcolor{blue}{For all systems listed, the lattice counts agree with $D_{\rm cont}=N_\phi$, and the quasihole manifolds are separated from higher states by positive finite-size gaps.}

\begin{table}[H]
\caption{Laughlin quasihole zero-mode counts and lowest positive energies after adding one flux quantum to the corresponding parent systems in Table~\ref{tab:supp_laughlin_counts}. Here $\nu_0=1/m$ is the parent filling and $N_\phi=mN+1$.}
\label{tab:supp_laughlin_qh}
\centering\small
\begingroup
\begin{ruledtabular}
\begin{tabular}{ccccccc}
$\nu_0$ & $N$ & $N_\phi$ & $L_x\times L_y$ & $D_{\rm lat}$ & $D_{\rm cont}$ & $\Delta_+/U$\\\hline
\multicolumn{7}{l}{\textbf{Fermions, $\nu_0=1/3$}}\\[2pt]
$1/3$ & 4 & 13 & $13\times 3$ & 13 & 13 & $9.32991\times10^{-2}$\\
$1/3$ & 5 & 16 & $8\times 6$ & 16 & 16 & $8.45115\times10^{-2}$\\
$1/3$ & 6 & 19 & $19\times 3$ & 19 & 19 & $9.32572\times10^{-2}$\\
$1/3$ & 7 & 22 & $11\times 6$ & 22 & 22 & $8.37661\times10^{-2}$\\
$1/3$ & 8 & 25 & $15\times 5$ & 25 & 25 & $8.01176\times10^{-2}$\\
$1/3$ & 9 & 28 & $12\times 7$ & 28 & 28 & $8.27506\times10^{-2}$\\
$1/3$ & 10 & 31 & $31\times 3$ & 31 & 31 & $9.33964\times10^{-2}$\\
\hline
\multicolumn{7}{l}{\textbf{Bosons, $\nu_0=1/4$}}\\[2pt]
$1/4$ & 4 & 17 & $17\times 3$ & 17 & 17 & $1.32466\times10^{-3}$\\
$1/4$ & 5 & 21 & $9\times 7$ & 21 & 21 & $6.43720\times10^{-3}$\\
$1/4$ & 6 & 25 & $15\times 5$ & 25 & 25 & $7.96950\times10^{-3}$\\
\end{tabular}
\end{ruledtabular}
\endgroup
\end{table}

For comparison, small systems at $\phi=1/2$ can have additional Laughlin zero modes. At $N=2$, the fermionic $\nu=1/3$ systems on $4\times3$ and $6\times2$ tori have four zero modes rather than three, while the bosonic $\nu=1/4$ systems on $4\times4$ and $8\times2$ tori have six rather than four. At the same flux density, the tested $N=3,4$ systems recover three fermionic or four bosonic zero modes: the fermionic geometries are $6\times3$ and $6\times4$, respectively; the bosonic ones are $6\times4$ and $8\times4$.

\subsection{\texorpdfstring{Moore--Read states on the torus}{Moore--Read states on the torus}}
\label{sec:supp_fermionic_MR14}
\label{sec:supp_general_k_kernel}
We next examine the Moore--Read (MR) states on the torus by diagonalizing their parent interactions in the torus LLL at flux density $\phi=1/2$. Table~\ref{tab:supp_torus_validation} compares the lattice zero-mode counts $D_{\rm lat}$ with the corresponding continuum counts $D_{\rm cont}$ and lists the finite-size gaps $\Delta_+$ above the zero-mode manifolds.

\emph{Models and diagnostics.---}We consider bosonic MR at $\nu=1/3$ and fermionic MR at $\nu=1/2,1/4$. The $\nu=1/2$ diagnostics are presented in the main text; here we give the explicit $\nu=1/4$ parent and its spectrum, correlations, and PES.

We consider the $(k,p)=(2,3)$ fermionic Moore--Read construction at $\nu=1/4$, with collective displacement $\bm\delta=\bm a_1$ and pair displacements $\{\bm\delta'_a\}=\{\bm0,\pm\bm a_2\}$. The on-site pair term vanishes by the Pauli principle. For $N_e=6$ fermions on an $8\bm a_1\times6\bm a_2$ torus at $\phi=1/2$, the parent interaction is:
\begin{equation}
H^{\rm fMR}_{1/4}
=
\sum_{\bm r}
\hat n_{\bm r}
\hat n_{\bm r+\bm a_1}
\hat n_{\bm r+2\bm a_1}
+
\frac12\sum_{\bm r}
\left(
\hat n_{\bm r} \hat n_{\bm r+\bm a_2}
+
\hat n_{\bm r} \hat n_{\bm r-\bm a_2}
\right).
\label{eq:supp_fMR14_H}
\end{equation}
The first term enforces a three-particle zero on each cluster $(\bm r,\bm r+\bm a_1,\bm r+2\bm a_1)$, while the second enforces pair zeros at separations $\pm\bm a_2$.

\begin{figure}[htbp]
    \centering
    \includegraphics[width=\textwidth]{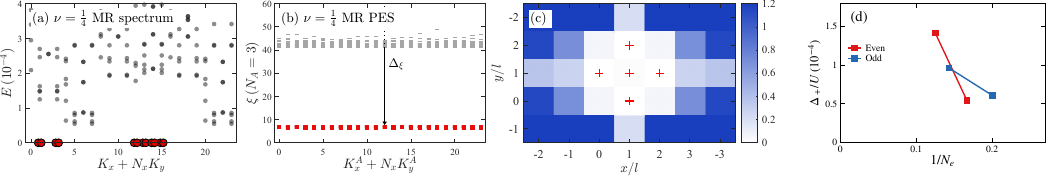}
    \caption{
    \textbf{Diagnostics of the fermionic $\nu=1/4$ Moore--Read state.}
    (a) Many-body spectrum for $N_e=6$ fermions on an $8\bm a_1\times 6\bm a_2$ torus with $\phi=1/2$, obtained from the Hamiltonian in Eq.~(\ref{eq:supp_fMR14_H}). 
    Red circles mark the twelve zero-energy states.
    (b) PES of the zero-energy manifold for $N_A=3$. 
    The arrow indicates the principal entanglement gap. The counting below the gap (1016) matches the expected quasihole counting of the continuum Moore--Read state.
    (c) Three-particle correlation function $g^{(3)}(\bm 0,-\bm r,\bm r)$. 
    Red crosses mark the locations associated with the imposed three-body and two-body clustering constraints.
    (d) Lowest positive energy $\Delta_+/U$ versus $1/N_e$ for the $8\bm a_1\times N_e\bm a_2$ tori in Table~\ref{tab:supp_torus_validation}. Red and blue squares denote even and odd $N_e$, respectively; lines connect the data points.
    }
    \label{fig:supp_fermionic_MR14}
\end{figure}

Figure~\ref{fig:supp_fermionic_MR14}(a) shows twelve zero modes, consistent with the expected torus degeneracy of the fermionic $\nu=1/4$ Moore--Read state. The PES for $N_A=3$ exhibits an infinite entanglement gap [Fig.~\ref{fig:supp_fermionic_MR14}(b)]. The three-particle correlation function $g^{(3)}(\bm0,-\bm r,\bm r)$ displays the clustering zeros imposed by the three-body and two-body interactions [Fig.~\ref{fig:supp_fermionic_MR14}(c)]. Figure~\ref{fig:supp_fermionic_MR14}(d) shows the finite-size dependence of the lowest positive energy, with even and odd particle numbers displayed separately.

\emph{Finite-size zero modes and gaps.---}The continuum MR parent at $\nu=1/(p+1)$ has torus zero-mode degeneracy~\cite{ArdonneTorusDegeneracy2008}
{
\begin{equation}
\label{eq:supp_continuum_torus_degeneracy}
D_{\rm cont}^{\rm MR}=(p+1)\begin{cases}3,&N\ \mathrm{even},\\1,&N\ \mathrm{odd},\end{cases}
\qquad N_\phi=(p+1)N.
\end{equation}
}
No quasiholes are added, and center-of-mass degeneracy is included.

The MR examples retain well-defined zero-mode manifolds across the sizes listed in Table~\ref{tab:supp_torus_validation}. Bosonic MR at $\nu=1/3$ has nine zero modes for $N=6,8$ and three for $N=5,7$, while fermionic MR at $\nu=1/2$ has six for $N=6, 8, 10, 12, 14$. Fermionic MR at $\nu=1/4$ shows a clear dependence on particle-number parity: the even-$N$ systems have twelve zero modes, whereas the odd-$N$ systems have four. The two sequences are therefore compared separately in Fig.~\ref{fig:supp_fermionic_MR14}(d). \textcolor{blue}{These counts agree with the corresponding continuum predictions, and the manifolds are separated from higher states by positive finite-size gaps.}

\begin{figure}[H]
\centering
\includegraphics[width=\textwidth]{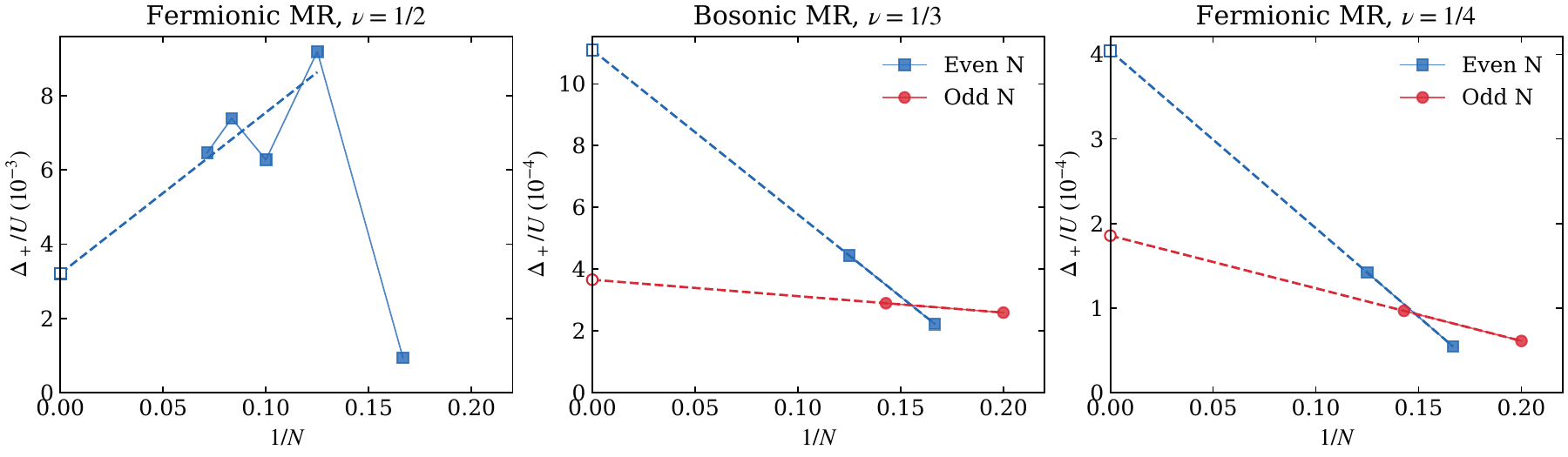}
\caption{\textbf{Finite-size scaling of the Moore--Read energy gaps.} The gap $\Delta_+/U$ above the zero-mode manifold is plotted against $1/N$, using the data and geometries in Table~\ref{tab:supp_torus_validation}. Solid lines connect data points; dashed lines are illustrative linear fits, $\Delta_+/U=\Delta_\infty/U+a/N$, with open symbols marking the intercepts. Left: fermions at $\nu=1/2$, fitted over $N=8,10,12,14$, giving $\Delta_\infty/U=0.00321$; including $N=6$ changes the intercept to $0.01170$. Middle: bosons at $\nu=1/3$. Right: fermions at $\nu=1/4$. The latter two panels fit the even-$N$ ($N=6,8$) and odd-$N$ ($N=5,7$) data separately; these two-point fits do not establish thermodynamic gaps.}
\label{fig:supp_mr_gap_scaling}
\end{figure}

Compared with the Laughlin examples, these MR systems exhibit larger finite-size fluctuations in the energy gap. The extrapolated gap depends sensitively on the system sizes included in the fit, while variations in the torus aspect ratio further complicate the finite-size scaling analysis. \textcolor{blue}{The positive fit intercepts at $\nu=1/2$ are suggestive of a nonzero thermodynamic gap [Fig.~\ref{fig:supp_mr_gap_scaling}], whereas the $\nu=1/3$ and $1/4$ data establish positive finite-size gaps only.}

\begin{table}[htbp]
\caption{Moore--Read parent zero-mode counts and lowest positive energies at $\phi=1/2$ and $U=1$. $\sigma=\mathrm B,\mathrm F$ denotes bosons and fermions, respectively; other notation follows Table~\ref{tab:supp_laughlin_counts}. Fermionic MR at $\nu=1/2$ uses the horizontal three-body density interaction of Fig.~\ref{fig:lattice_fMR}. Bosonic MR at $\nu=1/3$ adds the on-site term $U\sum_{\bm R}\hat n_{\bm R}(\hat n_{\bm R}-1)$; fermionic MR at $\nu=1/4$ instead adds $U\sum_{\bm R}\hat n_{\bm R}\hat n_{\bm R+\bm a_2}$.}
\label{tab:supp_torus_validation}
\centering\small
\begin{ruledtabular}
\begin{tabular}{cccccccccc}
$k$ & $\sigma$ & $\nu$ & $N$ & $N_\phi$ & $L_x\times L_y$ & $\bm\delta$ & $D_{\rm lat}$ & $D_{\rm cont}$ & $\Delta_+/U$\\\hline
\multicolumn{10}{l}{\textbf{Moore--Read: bosons, $\nu=1/3$}}\\[2pt]
2 & B & $1/3$ & 5 & 15 & $6\times5$ & $\bm a_1$ & 3 & 3 & $2.59097\times10^{-4}$ \\
2 & B & $1/3$ & 6 & 18 & $6\times6$ & $\bm a_1$ & 9 & 9 & $2.22224\times10^{-4}$ \\
2 & B & $1/3$ & 7 & 21 & $7\times6$ & $\bm a_1$ & 3 & 3 & $2.89556\times10^{-4}$ \\
2 & B & $1/3$ & 8 & 24 & $8\times6$ & $\bm a_1$ & 9 & 9 & $4.44113\times10^{-4}$ \\
\hline
\multicolumn{10}{l}{\textbf{Moore--Read: fermions, $\nu=1/2$}}\\[2pt]
2 & F & $1/2$ & 6 & 12 & $6\times4$ & $\bm a_1$ & 6 & 6 & $9.42163\times10^{-4}$ \\
2 & F & $1/2$ & 8 & 16 & $8\times4$ & $\bm a_1$ & 6 & 6 & $9.18931\times10^{-3}$ \\
2 & F & $1/2$ & 10 & 20 & $8\times5$ & $\bm a_1$ & 6 & 6 & $6.27293\times10^{-3}$ \\
2 & F & $1/2$ & 12 & 24 & $8\times6$ & $\bm a_1$ & 6 & 6 & $7.38991\times10^{-3}$ \\
2 & F & $1/2$ & 14 & 28 & $8\times7$ & $\bm a_1$ & 6 & 6 & $6.46589\times10^{-3}$ \\
\hline
\multicolumn{10}{l}{\textbf{Moore--Read: fermions, $\nu=1/4$}}\\[2pt]
2 & F & $1/4$ & 5 & 20 & $8\times5$ & $\bm a_1$ & 4 & 4 & $6.10963\times10^{-5}$ \\
2 & F & $1/4$ & 6 & 24 & $8\times6$ & $\bm a_1$ & 12 & 12 & $5.46658\times10^{-5}$ \\
2 & F & $1/4$ & 7 & 28 & $8\times7$ & $\bm a_1$ & 4 & 4 & $9.66762\times10^{-5}$ \\
2 & F & $1/4$ & 8 & 32 & $8\times8$ & $\bm a_1$ & 12 & 12 & $1.41967\times10^{-4}$ \\
\end{tabular}
\end{ruledtabular}
\end{table}

\emph{Continuum reference and direct quasihole counting.---}For MR states, adding $s$ flux quanta at fixed $N$ gives $N_\phi=(p+1)N+s$ and creates $2s$ elementary quasiholes. The continuum count includes all periodic admissible orbital configurations and all momentum sectors: at most two particles in any $2p+2$ consecutive orbitals and, for $p\ge1$, at most one particle in any $p$ consecutive orbitals.\footnote{B. A. Bernevig and N. Regnault, \href{https://doi.org/10.1103/PhysRevB.85.075128}{Phys. Rev. B \textbf{85}, 075128 (2012)}. The higher-$p$ root mapping is discussed in Ref.~\cite{ArdonneTorusDegeneracy2008}.}
We add one flux quantum to the lattice parent and compare its full zero-mode count with
{
\begin{equation}
\label{eq:supp_mr_one_flux_counting}
D_{\rm cont}^{\rm MR}=N_\phi\left(\left\lfloor\frac{N}{2}\right\rfloor+1\right),
\qquad N_\phi=(p+1)N+1.
\end{equation}
}
\textcolor{blue}{The MR systems in Table~\ref{tab:supp_mr_quasiholes} reproduce the continuum counts, with positive finite-size gaps above their quasihole manifolds.}

\begin{table}[H]
\caption{Moore--Read quasihole zero-mode counts and lowest positive energies after adding one flux quantum to the parent systems in Table~\ref{tab:supp_torus_validation}. Here $\nu_0$ is the parent filling and $N_\phi=N/\nu_0+1$.}
\label{tab:supp_mr_quasiholes}
\centering\small
\begingroup
\renewcommand{\arraystretch}{0.93}
\begin{ruledtabular}
\begin{tabular}{cccccccc}
$\sigma$ & $\nu_0$ & $N$ & $N_\phi$ & $L_x\times L_y$ & $D_{\rm lat}$ & $D_{\rm cont}$ & $\Delta_+/U$\\\hline
\multicolumn{8}{l}{\textbf{Moore--Read: bosons, $\nu_0=1/3$}}\\[2pt]
B & $1/3$ & 5 & 16 & $8\times4$ & 48 & 48 & $3.11872\times10^{-5}$\\
B & $1/3$ & 6 & 19 & $19\times2$ & 76 & 76 & $8.98152\times10^{-9}$\\
B & $1/3$ & 7 & 22 & $11\times4$ & 88 & 88 & $5.33952\times10^{-5}$\\
B & $1/3$ & 8 & 25 & $10\times5$ & 125 & 125 & $2.46982\times10^{-4}$\\\hline
\multicolumn{8}{l}{\textbf{Moore--Read: fermions, $\nu_0=1/2$}}\\[2pt]
F & $1/2$ & 6 & 13 & $13\times2$ & 52 & 52 & $8.33017\times10^{-5}$\\
F & $1/2$ & 8 & 17 & $17\times2$ & 85 & 85 & $1.66872\times10^{-4}$\\
F & $1/2$ & 10 & 21 & $7\times6$ & 126 & 126 & $3.15146\times10^{-3}$\\
F & $1/2$ & 12 & 25 & $10\times5$ & 175 & 175 & $5.54425\times10^{-3}$\\\hline
\multicolumn{8}{l}{\textbf{Moore--Read: fermions, $\nu_0=1/4$}}\\[2pt]
F & $1/4$ & 5 & 21 & $7\times6$ & 63 & 63 & $5.90427\times10^{-5}$\\
F & $1/4$ & 6 & 25 & $10\times5$ & 100 & 100 & $3.13944\times10^{-5}$\\
F & $1/4$ & 8 & 33 & $11\times6$ & 165 & 165 & $7.57855\times10^{-5}$\\
\end{tabular}
\end{ruledtabular}
\endgroup
\end{table}

For comparison, fermionic MR at $\nu=1/2$ and $N=4$ on a $4\times4$ torus has eight zero modes at zero boundary twist, rather than six. At $(\theta_x,\theta_y)/\pi=(0.17,0.29)$, the count returns to six. These complete-spectrum tests illustrate an accidental small-system degeneracy.

\subsection{\texorpdfstring{Fibonacci states on the torus}{Fibonacci states on the torus}}
\label{sec:supp_fibonacci_torus}
We consider the fermionic Fibonacci candidate at $\nu=3/5$ and $\phi=1/2$, using the four-body density interaction shown in Fig.~\ref{fig:lattice_fMR}. Tables~\ref{tab:supp_fibonacci_parent} and \ref{tab:supp_fibonacci_quasiholes} give the parent and quasihole results, respectively.

The continuum Fibonacci parent has~\cite{ArdonneTorusDegeneracy2008}
{
\begin{equation}
\label{eq:supp_fibonacci_parent_count}
D_{\rm cont}^{\rm Fib}=10,\qquad N_\phi=5N/3,\quad N\in3\mathbb Z.
\end{equation}
}
\emph{Finite-size zero modes and boundary conditions.---}The Fibonacci candidate at $\nu=3/5$ is more sensitive to torus geometry and boundary conditions. The $N=9$ and $N=15$ systems have the expected ten zero modes. At $N=12$, however, the $8\times5$ torus has twelve, while the $10\times4$ torus has ten. \textcolor{blue}{On the $8\times5$ torus, a small additional boundary twist $(\Delta\theta_x,\Delta\theta_y)/(2\pi)=(0.01,0)$ lifts the two extra zero modes, leaving the expected ten.} This sensitivity to a small boundary twist indicates that the extra zero modes reflect an accidental finite-size degeneracy rather than a topologically protected degeneracy.

\begin{table}[H]
\caption{Fibonacci parent zero-mode counts and lowest positive energies at $\nu=3/5$, $\phi=1/2$, and $U=1$, using the horizontal four-body density interaction of Fig.~\ref{fig:lattice_fMR}. Notation follows Tables~\ref{tab:supp_laughlin_counts} and \ref{tab:supp_torus_validation}.}
\label{tab:supp_fibonacci_parent}
\centering\small
\begin{ruledtabular}
\begin{tabular}{cccccccccc}
$k$ & $\sigma$ & $\nu$ & $N$ & $N_\phi$ & $L_x\times L_y$ & $\bm\delta$ & $D_{\rm lat}$ & $D_{\rm cont}$ & $\Delta_+/U$\\\hline
\multicolumn{10}{l}{\textbf{Fibonacci: fermions, $\nu=3/5$}}\\[2pt]
3 & F & $3/5$ & 9 & 15 & $6\times5$ & $\bm a_1$ & 10 & 10 & $3.97291\times10^{-4}$ \\
3 & F & $3/5$ & 12 & 20 & $8\times5$ & $\bm a_1$ & 12 & 10 & $1.04132\times10^{-4}$ \\
3 & F & $3/5$ & 12 & 20 & $10\times4$ & $\bm a_1$ & 10 & 10 & $1.28470\times10^{-3}$ \\
3 & F & $3/5$ & 15 & 25 & $10\times5$ & $\bm a_1$ & 10 & 10 & $9.37873\times10^{-4}$ \\
\end{tabular}
\end{ruledtabular}
\end{table}

\emph{Direct lattice quasihole counting.---}Adding $s$ flux quanta produces $3s$ elementary quasiholes. The continuum reference counts periodic orbital configurations with at most three particles in any five consecutive orbitals, including all translations and momentum sectors. For one added flux quantum,
{
\begin{equation}
\label{eq:supp_fibonacci_one_flux_counting}
D_{\rm cont}^{\rm Fib}=\frac{N_\phi}{3}\left(\frac{N}{3}+1\right)\left(\frac{2N}{3}+3\right),
\qquad N_\phi=\frac{5N}{3}+1,\quad N\in3\mathbb Z.
\end{equation}
}
The direct lattice counts are listed in Table~\ref{tab:supp_fibonacci_quasiholes}.

\textcolor{blue}{The $N=9$ Fibonacci system on the $8\times4$ torus has $256$ zero modes, exceeding the continuum count of $192$. The $N=12$ systems on the $7\times6$ torus have $385$ zero modes for both tested boundary conditions, matching the continuum count. This size comparison supports a finite-size interpretation of the excess modes, with geometry and boundary conditions also affecting the results.}

The available Fibonacci sizes and their sensitivity to geometry and boundary conditions do not establish a thermodynamic gap.

\begin{table}[H]
\caption{Fibonacci quasihole zero-mode counts and lowest positive energies after adding one flux quantum at fixed particle number, $N_\phi=5N/3+1$. The $8\times4$ system uses $(\theta_x,\theta_y)/\pi=(1,1)$. The two $7\times6$ systems use physical boundary twists $(\theta_x,\theta_y)/\pi=(0,1)$ ($\mathrm a$) and $(1,0)$ ($\mathrm b$).}
\label{tab:supp_fibonacci_quasiholes}
\centering\small
\begingroup
\begin{ruledtabular}
\begin{tabular}{cccccccc}
$\sigma$ & $\nu_0$ & $N$ & $N_\phi$ & $L_x\times L_y$ & $D_{\rm lat}$ & $D_{\rm cont}$ & $\Delta_+/U$\\\hline
\multicolumn{8}{l}{\textbf{Fibonacci: fermions, $\nu_0=3/5$}}\\[2pt]
F & $3/5$ & 9 & 16 & $8\times4$ & 256 & 192 & $3.06649\times10^{-5}$\\
F & $3/5$ & 12 & 21 & $7\times6^{\mathrm a}$ & 385 & 385 & $1.84551\times10^{-4}$\\
F & $3/5$ & 12 & 21 & $7\times6^{\mathrm b}$ & 385 & 385 & $1.85707\times10^{-4}$\\
\end{tabular}
\end{ruledtabular}
\endgroup
\end{table}

\subsection{\texorpdfstring{$C_4$-symmetric nearest-neighbor interactions}{C4-symmetric nearest-neighbor interactions}}
\label{sec:supp_c4_nn}
\begin{figure}[htbp]
    \centering
    \includegraphics[width=0.50\textwidth]{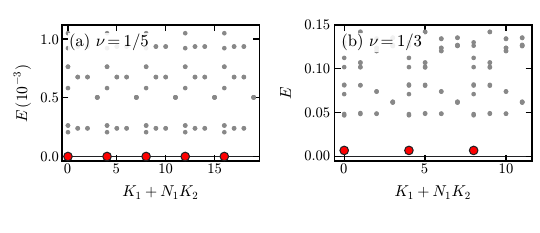}
    \caption{
    \textbf{Many-body spectra of the $C_4$-symmetric nearest-neighbor interaction.}
    Spectra of Eq.~\eqref{eq:supp_c4_nn} for $N_e=4$ fermions at fixed magnetic flux density $\phi=1/5$.
    (a) At $\nu=\phi=1/5$, with $N_\phi=20$ on a $10\bm a_1\times10\bm a_2$ lattice, five exact zero modes are separated from the excited states by a finite-size gap $\Delta=2.07\times10^{-4}$.
    (b) At $\nu=1/3$, with $N_\phi=12$ on a $10\bm a_1\times6\bm a_2$ lattice, the three lowest states have minimum energy $E_0=0.00690232$ and splitting $6.67\times10^{-6}$; no zero modes are present.
    }
    \label{fig:supp_c4_nn}
\end{figure}

We now ask whether the finite-displacement construction can yield model FQH states compatible with the symmetry of the underlying lattice. On the square lattice, we therefore consider a $C_4$-symmetric parent interaction with equal nearest-neighbor repulsions along $\bm a_1$ and $\bm a_2$. With $U=1$, the LLL-projected Hamiltonian used for exact diagonalization is
\begin{equation}
    \hat H_{\rm NN} = \mathcal P_{\mathrm{LLL}}\sum_{\bm r}
    \left(\hat n_{\bm r}\hat n_{\bm r+\bm a_1}
    +\hat n_{\bm r}\hat n_{\bm r+\bm a_2}
    \right)\mathcal P_{\mathrm{LLL}}. \label{eq:supp_c4_nn}
\end{equation}
Since the density-interaction terms are positive semidefinite, a zero-energy wavefunction must vanish whenever two particles are separated by $\pm\bm a_1$ or $\pm\bm a_2$. Together with the contact zero required by fermionic antisymmetry, these conditions require five distinct pair zeros. They are realized by the $m=5$, $\nu=1/5$ state in Eq.~\eqref{eq:laughlin}, with $\{\bm\delta_l\}=\{\bm0,\pm\bm a_1,\pm\bm a_2\}$. This state is therefore an exact zero mode. Consistently, exact diagonalization for $N_e=4$ at $\nu=\phi=1/5$ yields five zero modes separated from the excited states by a finite-size gap $\Delta=2.07\times10^{-4}$ [Fig.~\ref{fig:supp_c4_nn}(a)].

The additional constraints along $\bm a_2$ also affect the filling at which zero modes can occur. In the pair-factor construction, the contact zero and the two zeros at $\pm\bm a_1$ give the $m=3$, $\nu=1/3$ state. Requiring zeros at $\pm\bm a_2$ increases the minimum pair-factor degree from three to five, corresponding to $\nu=1/5$. Keeping $N_e=4$ and $\phi=1/5$ fixed, we next examine $\nu=1/3$ on a $10\bm a_1\times6\bm a_2$ torus. The lowest energy is $E_0=0.00690232>0$, so the same interaction has no exact zero modes in this system [Fig.~\ref{fig:supp_c4_nn}(b)]. \textcolor{blue}{Thus, within this construction, the zero requirements imposed by interaction symmetry constrain the compatible filling.}

\subsection{Small-flux continuum limit and restoration of rotational symmetry}
\label{sec:supp_continuum_limit}
{
The preceding subsection imposes lattice symmetry by adding interaction directions at finite flux. Here we keep a single interaction direction and examine the small-flux limit, in which the displacement vanishes in magnetic-length units and continuum contact clustering is recovered. We illustrate this limit using the fermionic $\nu=1/3$ state and its directional nearest-neighbor parent.

As in the main text, we set $\ell_B=1$ and measure all lengths in magnetic-length units. For a square lattice with spacing $a=|\bm a_1|=|\bm a_2|$, the flux density and complex coordinates are
\begin{equation}
\label{eq:supp_continuum_coordinates}
 \phi=\frac{a^2}{2\pi},\qquad
 z_i=\frac{x_i+iy_i}{\sqrt2},\qquad
 \delta=\frac{a\,e^{i\theta}}{\sqrt2}
       =\sqrt{\pi\phi}\,e^{i\theta},
\end{equation}
where $\bm r_i=(x_i,y_i)$ and $\delta$ is the complex coordinate of the displacement $\bm\delta=a\hat{\bm e}_\theta$, with $\hat{\bm e}_\theta=(\cos\theta,\sin\theta)$. We take $\phi\to0$ at fixed $N$, $N_\phi$, and torus geometry in magnetic-length units, so that $\delta\to0$ while the number of lattice sites $N_{\rm sites}=N_\phi/\phi$ increases.

\emph{Wavefunction.---}For the displacements $\{0,\pm\delta\}$, write $z_{ij}=z_i-z_j$ and $\Psi_\delta\equiv\Psi_{\{0,\pm\delta\}}^{\nu=1/3}$. Omitting the common Gaussian factor, the planar polynomial and its expansion for small $\delta$ are
\begin{equation}
\label{eq:supp_continuum_wavefunction}
 \Psi_\delta(\bm z)=\prod_{i<j}z_{ij}(z_{ij}^2-\delta^2),\qquad
 \Psi_\delta=\Psi_{\rm L}
 -\delta^2\sum_{i<j}z_{ij}\!\prod_{\substack{k<l\\(k,l)\ne(i,j)}}z_{kl}^3
 +O(\delta^4),\qquad
 \Psi_{\rm L}=\prod_{i<j}z_{ij}^3.
\end{equation}
The leading correction carries the directional phase $e^{2i\theta}$ and has amplitude $O(\phi)$ at fixed particle number. As $\delta\to0$, the three pair zeros merge into a cubic contact zero, and the normalized state approaches the continuum Laughlin state.

\emph{Interaction.---}We next examine the corresponding parent interaction, whose microscopic bonds remain directional at nonzero $\phi$. Following Eq.~\eqref{parentH}, we include both $\pm\bm\delta$ and write
\begin{equation}
\label{eq:supp_continuum_lattice_parent}
 H_\delta=U\sum_{\bm R}\hat n_{\bm R}
 \left(\hat n_{\bm R+\bm\delta}+\hat n_{\bm R-\bm\delta}\right).
\end{equation}
The Kapit--Mueller lowest band consists of sampled continuum LLL wavefunctions~\cite{Kapit_Exact_2010}, allowing us to express this interaction in terms of the continuum density as $a\to0$. Since each lattice site represents an area $a^2$, we have $\hat n_{\bm R}\simeq a^2\hat n(\bm R)$ and $a^2\sum_{\bm R}f(\bm R)\to\int d^2r\,f(\bm r)$. The two density factors contribute $a^4$, while the lattice sum contributes $a^{-2}$, giving $H_\delta\simeq Ua^2\int d^2r\,:\hat n(\bm r)[\hat n(\bm r+\bm\delta)+\hat n(\bm r-\bm\delta)]:$. Rewriting this result in the standard two-particle form gives
\begin{equation}
\label{eq:supp_continuum_pair_potential}
\begin{aligned}
 H_\delta&\simeq\frac12\int d^2r_1d^2r_2\,
 V_a(\bm r_1-\bm r_2):\hat n(\bm r_1)\hat n(\bm r_2):,\\
 V_a(\bm r)&=2Ua^2\left[\delta^{(2)}(\bm r-\bm\delta)
                  +\delta^{(2)}(\bm r+\bm\delta)\right].
\end{aligned}
\end{equation}

Expanding the displaced delta functions in powers of $a$ gives
\begin{equation}
\label{eq:supp_continuum_potential_expansion}
 V_a=4Ua^2\delta^{(2)}(\bm r)
    +2Ua^4(\hat{\bm e}_\theta\!\cdot\!\bm\nabla)^2\delta^{(2)}(\bm r)
    +\frac{Ua^6}{6}(\hat{\bm e}_\theta\!\cdot\!\bm\nabla)^4\delta^{(2)}(\bm r)+\cdots.
\end{equation}
The contact term has no effect on fermionic states. Within the fermionic LLL, the second-derivative term has matrix elements independent of the bond direction $\theta$. To show this, consider two arbitrary fermionic LLL wavefunctions near coincidence in the relative coordinate $z_{12}=(r_x+ir_y)/\sqrt2$. Fermionic antisymmetry under $z_{12}\to-z_{12}$ restricts their analytic factors to odd powers:
\begin{equation}
\label{eq:supp_continuum_relative_states}
 \psi(z_{12})=\left(c_1z_{12}+c_3z_{12}^3+\cdots\right)e^{-|z_{12}|^2/4},\qquad
 \chi(z_{12})=\left(d_1z_{12}+d_3z_{12}^3+\cdots\right)e^{-|z_{12}|^2/4}.
\end{equation}
Their product is $\psi^*\chi=\frac12c_1^*d_1(r_x^2+r_y^2)+O(|\bm r|^4)$. At coincidence, its second derivative along any direction therefore equals half its Laplacian. Integrating by parts twice then gives
\begin{equation}
\label{eq:supp_continuum_derivative_identity}
 \langle\psi|(\hat{\bm e}_\theta\!\cdot\!\bm\nabla)^2\delta^{(2)}(\bm r)|\chi\rangle
 =\frac12\langle\psi|\nabla^2\delta^{(2)}(\bm r)|\chi\rangle
 =c_1^*d_1.
\end{equation}
Thus, within the fermionic LLL, the leading nonzero interaction is equivalent to $Ua^4\mathcal P_{\mathrm{LLL}}\nabla^2\delta^{(2)}\mathcal P_{\mathrm{LLL}}$, where $\mathcal P_{\mathrm{LLL}}$ denotes LLL projection. This is proportional to the rotationally invariant $V_1$ pseudopotential~\cite{TrugmanKivelson85,haldane_fractional_1983}; its independence of the bond direction follows from fermionic antisymmetry and LLL holomorphicity.

\textcolor{blue}{As $\phi\to0$, the model wavefunction and the normalized projected interaction therefore approach their continuum Laughlin and $V_1$ forms together.}
}

\subsection{Robustness under short-range hopping}
\label{sec:supp_truncation_fci}

\begin{figure}[H]
    \centering
    \includegraphics[width=6.40in]{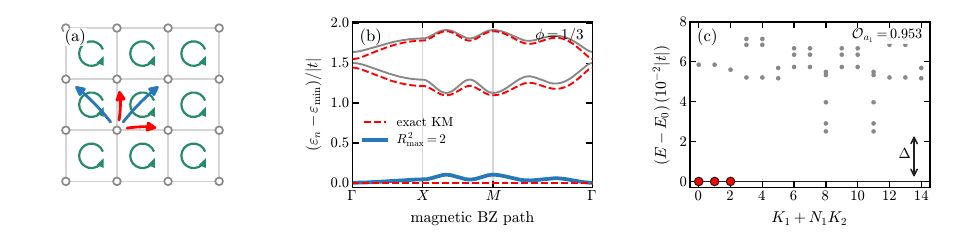}
    \caption{\textbf{Hopping truncation for the fermionic $\nu=1/3$ Laughlin state at $\phi=1/3$.}
    (a) Nearest-neighbor (red) and diagonal (blue) hopping retained at $R_{\max}^2=2$. Green arrows indicate the magnetic flux.
    (b) Single-particle spectra of the untruncated (red dashed) and truncated (solid) Kapit--Mueller models. The truncated lowest $C=1$ band is blue, and the upper bands are gray. Each spectrum is measured relative to its own lowest-band minimum.
    (c) Many-body spectrum of Eq.~\eqref{eq:supp_truncated_km}, obtained without band projection for $N_e=5$ on a $9\times5$ torus at $U=5$. Energies are measured relative to the ground-state energy $E_0$. Red circles mark the three nearly degenerate lowest states, and the arrow indicates the gap $\Delta$ above them. Their subspace overlap with the exact Laughlin states is $\mathcal O_{a_1}=0.953$.
    }
    \label{fig:supp_truncation_fci}
\end{figure}

Our exact model states are constructed in the flat lowest band of the Kapit--Mueller Hamiltonian, which contains long-range hopping. To test the robustness of the Laughlin state against hopping truncation, we truncate the hopping in Eq.~\eqref{eq:KM} before imposing torus periodicity, while keeping the nearest-neighbor interaction $\hat V_{a_1}=\sum_{\bm r}\hat n_{\bm r}\hat n_{\bm r+\bm a_1}$ unchanged. At $\nu=1/3$, the resulting Hamiltonian is
{\begin{equation}
 \hat H_{a_1}^{(R_{\max})}
 =\hat H_{\mathrm{KM}}^{(R_{\max})}+U\hat V_{a_1}.
 \label{eq:supp_truncated_km}
\end{equation}}
Here $\hat H_{\mathrm{KM}}^{(R_{\max})}$ retains only terms with $m^2+n^2\le R_{\max}^2$ in Eq.~\eqref{eq:KM}, where $\bm d=m\bm a_1+n\bm a_2$.
We choose $R_{\max}^2=2$, retaining only nearest-neighbor and diagonal hopping [Fig.~\ref{fig:supp_truncation_fci}(a)] and yielding a short-range extended Harper--Hofstadter model at uniform flux $\phi=1/3$~\cite{Kapit_Exact_2010}. Figure~\ref{fig:supp_truncation_fci}(b) compares the single-particle spectra of the truncated and untruncated Kapit--Mueller models along the path $\Gamma$--$X$--$M$--$\Gamma$ in the magnetic Brillouin zone. After truncation, the lowest band retains Chern number $C=1$. On the $9 \times 5$ torus used below, its bandwidth $W/|t|=0.0871$ is much smaller than the indirect gap $\Delta_{\rm band}/|t|=1.03$ to the upper bands. On the $9\times5$ torus used below, the lowest-band subspaces of the truncated and untruncated models nearly coincide, with a minimum squared principal overlap $\mathcal O_{\min}^{(1)}=0.999$.

We next diagonalize the full Hamiltonian in Eq.~\eqref{eq:supp_truncated_km}, without band projection, for $N_e=5$ fermions on a $9\times5$ torus with $N_\phi=15$, corresponding to $\nu=\phi=1/3$. At $U=5$, the three lowest states occupy momentum sectors $(K_1,K_2)=(0,0),(1,0),(2,0)$. \textcolor{blue}{These states form a nearly degenerate manifold with splitting $\delta E=E_3-E_1=4.43\times10^{-5}|t|$, well below the gap $\Delta=E_4-E_3=0.0251|t|$ to the remaining spectrum [Fig.~\ref{fig:supp_truncation_fci}(c)].}

To quantify the connection to the exact Laughlin states, we compare the corresponding many-body subspaces. Let $P_{a_1}^{(R_{\max})}$ project onto the three lowest states of Eq.~\eqref{eq:supp_truncated_km}, and $P_{\rm L}$ onto the three exact Laughlin zero modes in the untruncated Kapit--Mueller lowest band on the same torus.  Both projectors are represented in the same site-occupation basis.  \textcolor{blue}{Their subspace overlap is $\mathcal O_{a_1}=\frac13\operatorname{Tr}\!\left(P_{a_1}^{(R_{\max})}P_{\rm L}\right)=0.953$.} This comparison accounts for both the change in the lowest-band subspace and upper-band admixture.  Averaged over the three lowest states, the upper-band occupation per particle is $\bar n_{\rm upper}=(3N_e)^{-1}\sum_{i=1}^{3}\langle\Psi_i|\hat N_{\rm upper}^{\rm tr}|\Psi_i\rangle=8.10\times10^{-4}$, where $\hat N_{\rm upper}^{\rm tr}$ counts particles in the upper bands of the truncated model. Thus, even without band projection, the three low-energy states remain predominantly within the lowest band of the truncated model. \textcolor{blue}{Together, these results support the robustness of the Laughlin state against this hopping truncation in the finite system studied.}

\FloatBarrier

\section{Interaction-range-induced phase transitions}
\label{sec:supp_phase_diagram_ne8}

For completeness, we provide the exact-diagonalization data underlying the interaction-range phase diagram discussed in the main text. 
The calculations are performed for $N_e=8$ fermions at filling $\nu=1/3$. 
The horizontal axis is the interaction displacement in units of the lattice constant, and the vertical axis is the magnetic flux density $\phi$. 
Each computed point is classified using combined diagnostics from the many-body spectrum, PES, and projected structure factor.

The FQH-like region is identified by the persistence of the Laughlin-type low-energy structure: the spectrum contains the expected low-energy manifold, the PES has the appropriate low-lying counting below a principal entanglement cutoff, and the projected structure factor does not show strong finite-momentum peaks. 
Points labeled as non-FQH fail one or more of these diagnostics, for example by developing an enlarged low-energy manifold, anomalous PES counting, or a pronounced density-modulation peak in $\bar S(\bm q)$. 
The gray region denotes parameter points that were not reached.

\begin{figure}[H]
    \centering
    \includegraphics[width=0.86\textwidth]{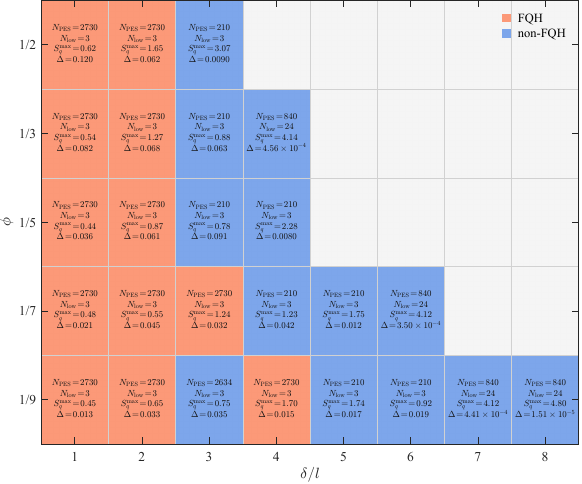}
    \caption{
    \textbf{Exact-diagonalization data for the $N_e=8$ phase diagram.}
    Phase diagram at $\nu=1/3$ as a function of interaction displacement $|\bm\delta|/l$ and flux density $\phi$, where $l$ is the lattice spacing. 
    The warm-colored region denotes points identified as FQH-like, the blue region denotes non-FQH behavior, and the gray region denotes points without numerical data. 
    The annotations inside each computed cell show, from top to bottom, the PES counting $N_{\rm PES}$, the number of low-energy states $N_{\rm low}$, the maximal projected structure factor $\bar S_q^{\max}$, and the many-body gap $\Delta$.
    }
    \label{fig:supp_phase_diagram_ne8}
\end{figure}

In Fig.~\ref{fig:supp_phase_diagram_ne8}, $N_{\rm PES}$ denotes the number of PES levels below the chosen cutoff $\xi_{\rm cut}$, $N_{\rm low}$ denotes the number of low-energy states identified from the many-body spectrum, $\bar S_q^{\max}$ is the maximum of the projected structure factor, and $\Delta$ is the many-body gap above the zero-energy manifold. 
The values of $\xi_{\rm cut}$ are chosen point by point from the visible principal gap in the corresponding PES.

\FloatBarrier

\section{Macdonald-polynomial construction} \label{sec:supp_recursive_mr}
In this section of the SM, we explain the Macdonald polynomial and its relationship to our planar lattice model wavefunction in more detail. Our discussion follows the Jack-polynomial description of continuum FQH wavefunctions developed by Andrei Bernevig and Duncan Haldane~\cite{Bernevig_Jack08}. The key point is that the lattice-specific clustering property is encoded in the Macdonald polynomial for the entire Read--Rezayi $Z_k$ series.

We organize this section as follows. We start with a toy example to illustrate how Jack and Macdonald polynomials work by explicitly deriving the four-particle Moore--Read wavefunction from its root through the Jack and Macdonald eigenvalue equations and verifying its finite-displacement clustering. Next, we discuss the general route of obtaining lattice model wavefunctions from Macdonald polynomials for the entire Read--Rezayi sequence. 

\subsection{Toy example: the four-particle Moore--Read construction}
\label{sec:supp_root_derivation}
In the continuum, Jack polynomial is determined from its root and the Laplace--Beltrami eigenvalue equation. Similarly, the Macdonald polynomial is determined by the same root but with a difference eigenvalue equation. In this section, we start with the simple yet nontrivial example of the four-particle Moore-Read wavefunction to illustrate how to systematically obtain the Jack and Macdonald polynomials. Through the concrete calculation, the continuum and lattice clustering rules are manifestly seen.

Our concrete goal is to construct a symmetric four-particle Moore--Read factor which vanishes identically whenever three particles occupy $z_0,z_0+\delta,z_0+2\delta$, independently of the fourth particle. In the $\delta = 0$ case, the continuum limit is recovered. The starting point of the polynomial construction is a root configuration which physically reflects the generalized Pauli principle: Minimum-degree bosonic states follow the $(k,2)$-admissible root, which means there are no more than $k$ particles in any two consecutive orbitals. For Moore--Read states, $k=2$. In our four-particle study, the root configuration is hence 
$\lambda=(2,2,0,0)$, with occupation pattern $202$. The Jack and Macdonald polynomials for continuum and lattice wavefunctions share the same root configuration but their eigenvalue equations are different.

Squeezing the root $\lambda$ gives two additional partitions, $(2,1,1,0)$ and $(1,1,1,1)$. Let $m_\mu$ denote the sum over distinct permutations of $x_1^{\mu_1}\cdots x_4^{\mu_4}$. In monic normalization, the coefficient of the root monomial is unity, so the expansion is
\begin{equation}
\label{eq:supp_mr4_monomial_basis}
\begin{gathered}
    F=m_{22}+A\,m_{211}+B\,m_{1111},\\
    m_{22}=\sum_{i<j}x_i^2x_j^2,\qquad
    m_{211}=\sum_{i=1}^4\ \sum_{\substack{j<\ell\\j,\ell\ne i}}x_i^2x_jx_\ell,\qquad
    m_{1111}=x_1x_2x_3x_4.
\end{gathered}
\end{equation}
Trailing zeros in the subscripts are omitted. The Jack and Macdonald operators act on this same three-dimensional space and determine the two coefficients $A$ and $B$.

\emph{Jack polynomial.---} The Jack polynomial satisfies the Laplace--Beltrami eigenvalue equation~\cite{Bernevig_Jack08}, $\mathcal H_{\rm J}^{(\alpha)}J_\lambda^{(\alpha)}=e_\lambda J_\lambda^{(\alpha)}$, where
\begin{equation}
\label{eq:supp_mr4_jack_operator}
 \mathcal H_{\rm J}^{(\alpha)}
 =\sum_i(x_i\partial_{x_i})^2
 +\frac{1}{\alpha}\sum_{i<j}\frac{x_i+x_j}{x_i-x_j}
 (x_i\partial_{x_i}-x_j\partial_{x_j}),\qquad
 e_\lambda=\sum_{i=1}^N\left[\lambda_i^2+\frac{N+1-2i}{\alpha}\lambda_i\right].
\end{equation}
Acting on the three basis polynomials gives
\begin{equation}
\label{eq:supp_mr4_jack_action}
\begin{aligned}
 \mathcal H_{\rm J}^{(\alpha)}m_{22}
   &=(8+8/\alpha)m_{22}+(4/\alpha)m_{211},\\
 \mathcal H_{\rm J}^{(\alpha)}m_{211}
   &=(6+6/\alpha)m_{211}+(24/\alpha)m_{1111},\\
 \mathcal H_{\rm J}^{(\alpha)}m_{1111}&=4m_{1111}.
\end{aligned}
\end{equation}
Substituting $F$ into the eigenvalue equation and matching the coefficients of $m_{211}$ and $m_{1111}$ gives
\begin{equation}
\label{eq:supp_mr4_jack_coefficients}
 (e_{22}-e_{211})A=\frac4\alpha,\qquad
 (e_{22}-e_{1111})B=\frac{24}{\alpha}A,
 \qquad
 A=\frac{2}{\alpha+1},\quad B=\frac{12}{(\alpha+1)(\alpha+2)}.
\end{equation}
At the Moore--Read parameter $\alpha=-3$, this gives
\begin{equation}
\label{eq:supp_mr4_contact_polynomial}
\begin{aligned}
 J_{22}^{(-3)}(\bm x)&=m_{22}-m_{211}+6m_{1111}\\
 &=\frac12\left[(x_1-x_2)^2(x_3-x_4)^2
 +(x_1-x_3)^2(x_2-x_4)^2+(x_1-x_4)^2(x_2-x_3)^2\right].
\end{aligned}
\end{equation}
The pair-factor form makes the three-particle contact zero explicit: every term vanishes when any three coordinates coincide.

\emph{Macdonald polynomial.---}To deform this contact zero, we replace the differential eigenvalue problem by the Macdonald difference equation. For $N$ variables ($N=4$ here), the operator and its eigenvalue are~\cite{FeiginMacdonaldWheel2003}

\begin{equation}
    \mathcal D_{\rm M}
    =\sum_i\prod_{j\ne i}\frac{t_{\rm M}x_i-x_j}{x_i-x_j}\,T_{q_{\rm M},x_i},
    \qquad
    \mathcal D_{\rm M}P_\lambda
    =\left(\sum_iq_{\rm M}^{\lambda_i}t_{\rm M}^{N-i}\right)P_\lambda,
    \label{eq:macdonald_eigenproblem}
\end{equation}

Here $T_{q_{\rm M},x_i}$ replaces $x_i$ by $q_{\rm M}x_i$. For brevity, we write $q=q_{\rm M}$ and $t=t_{\rm M}$. In this monomial basis, $E_\lambda$ denotes a diagonal coefficient and $c_{ij}$ an off-diagonal mixing coefficient. Since the operator is triangular, the $E_\lambda$ are also its eigenvalues, rather than physical energies. Explicitly,
\begin{equation}
\label{eq:supp_mr4_macdonald_action}
\begin{aligned}
 \mathcal D_{\rm M}m_{22}&=E_{22}m_{22}+c_{10}m_{211}+c_{20}m_{1111},\\
 \mathcal D_{\rm M}m_{211}&=E_{211}m_{211}+c_{21}m_{1111},\\
 \mathcal D_{\rm M}m_{1111}&=E_{1111}m_{1111},
\end{aligned}
\end{equation}
where
\begin{equation}
\label{eq:supp_mr4_macdonald_matrix}
\begin{gathered}
 E_{22}=(t+1)(q^2t^2+1),\qquad
 E_{211}=(qt+1)(qt^2+1),\qquad
 E_{1111}=q(t+1)(t^2+1),\\
 c_{10}=t(q^2-1)(t-1),\qquad
 c_{20}=-(q^2-1)(t-1)^3,\\
 c_{21}=(q-1)(t-1)\left[3qt^2+2qt+q+t^2+2t+3\right].
\end{gathered}
\end{equation}
Matching coefficients in $\mathcal D_{\rm M}F=E_{22}F$ now gives
\begin{equation}
\label{eq:supp_mr4_macdonald_coefficients}
 A=\frac{c_{10}}{E_{22}-E_{211}}
 =\frac{(q+1)(t-1)}{qt-1},\qquad
 B=\frac{c_{20}+A c_{21}}{E_{22}-E_{1111}}
 =\frac{(q+1)(t-1)^2(2qt+q+t+2)}{(qt-1)(qt^2-1)}.
\end{equation}
These coefficients determine the monic polynomial $P_{22}(\bm x;q,t)$. Along $q=t^\alpha$, their $t\to1$ limits recover the Jack coefficients above. We obtain the clustering polynomial by regular specialization of these generic-parameter expressions~\cite{FeiginMacdonaldWheel2003}.

For the Moore--Read wheel condition, we set $q=t^{-3}$ and define $\widehat P_{22}(\bm x;t)=P_{22}(\bm x;t^{-3},t)$, obtaining
\begin{equation}
    \widehat P_{22}=m_{22}+A(t)m_{211}+B(t)m_{1111}, \label{eq:supp_mr4_wheel_polynomial}
\end{equation}
where
\begin{equation}
    A(t)=1-t-t^{-1},\qquad B(t)=\frac{(t^2-t+1)(t^4+2t^3+2t+1)}{t^3}.
\end{equation}

It is important to notice that $\widehat P_{22}(x,tx,t^2x,y;t)=0$ for arbitrary $x$ and $y$. This is the so-called wheel condition satisfied by the Macdonald polynomials. As $t\to1$, $A(t)\to-1$ and $B(t)\to6$, recovering the contact Jack polynomial.

\emph{Finite-displacement polynomial.---}We now convert the multiplicative wheel into the required zero at finite separation. Define $\xi_i=z_i-\frac14\sum_jz_j$ and set $x_i=e^{\epsilon\xi_i}$ and $t=e^{\epsilon\delta}$. A particle separation $z_j-z_i=\delta$ then gives $x_j=tx_i$, so the configuration $z_0,z_0+\delta,z_0+2\delta$ maps to $x,tx,t^2x$. At fixed $\delta$, the coefficient functions expand as
\begin{equation}
\label{eq:supp_mr4_coefficient_expansion}
\begin{aligned}
 A(e^{\epsilon\delta})&=-1-\epsilon^2\delta^2-\frac{\epsilon^4\delta^4}{12}+O(\epsilon^6),\\
 B(e^{\epsilon\delta})&=6+12\epsilon^2\delta^2+8\epsilon^4\delta^4+O(\epsilon^6).
\end{aligned}
\end{equation}
We expand each monomial as $\bm x^\mu=\exp(\epsilon\sum_i\mu_i\xi_i)$ and sum over distinct permutations. All terms below order $\epsilon^4$ cancel. With $z_{ij}=z_i-z_j$, the leading term is
\begin{equation}
\label{eq:supp_mr4_leading_term}
 \widehat P_{22}(e^{\epsilon\bm\xi};e^{\epsilon\delta})
 =\epsilon^4\left[
 J_{22}^{(-3)}(\bm z)-\delta^2\sum_{i<j}z_{ij}^2+7\delta^4
 \right]+O(\epsilon^5).
\end{equation}
The coefficient of $\epsilon^4$ defines the finite-displacement clustering polynomial,
\begin{equation}\color{blue}
    \mathcal W_{4,\delta}^{(2)}(\bm z) = \frac12\bigl[z_{12}^{2}z_{34}^{2}+z_{13}^{2}z_{24}^{2}+z_{14}^{2}z_{23}^{2}\bigr] - \delta^2 \sum_{i<j}z_{ij}^2+7\delta^4. \label{eq:supp_n4_explicit}
\end{equation}
This polynomial is symmetric and translation invariant, and at $\delta=0$ it reduces to the monic bosonic Moore--Read polynomial. Its finite-displacement zero follows directly by fixing two particles at $z_0$ and $z_0+\delta$:
\begin{equation}
    \mathcal W_{4,\delta}^{(2)}(z_0,z_0+\delta,z_3,z_4) = \prod_{a=3}^{4} (z_a-z_0+\delta)(z_a-z_0-2\delta). \label{eq:supp_n4_pair_pin}
\end{equation}
Setting either remaining particle at $z_0+2\delta$ makes one factor vanish, independently of the other particle. Permutation symmetry then establishes the clustering condition in Eq.~\eqref{eq:wheel} for any triplet. The Macdonald deformation thus replaces the Moore--Read contact zero by the prescribed zero at finite separation.

\subsection{General Read--Rezayi construction}  \label{sec:supp_general_macdonald}
Here we discuss how to use Macdonald polynomials to construct lattice bosonic $Z_k$ Read--Rezayi wavefunctions. For $N=kM$ particles, the bosonic Read--Rezayi root occupation pattern $k0k0\cdots k$ corresponds to the partition,
\begin{equation}
    \lambda_k^{(N)}
    =([2(M-1)]^k,[2(M-2)]^k,\ldots,2^k,0^k),
    \qquad
    D_{N,k}=|\lambda_k^{(N)}|=\frac{N(N-k)}{k},
    \label{eq:supp_rr_root}
\end{equation}
Here $[a]^k$ denotes $k$ repetitions of $a$. The generalized Pauli principle permits at most $k$ particles in any two consecutive orbitals, corresponding to the $(k,2)$ admissibility condition $\lambda_i-\lambda_{i+k}\ge2$. The partition in Eq.~\eqref{eq:supp_rr_root} has the minimum weight among admissible partitions with $N$ particles, fixing the total degree $D_{N,k}$ of the corresponding contact-state Jack polynomial $J_{\lambda_k^{(N)}}^{(-(k+1))}$~\cite{Bernevig_Jack08}.

Using the monomial basis defined above, the monic Macdonald polynomial has the triangular expansion $P_\lambda=m_\lambda+\sum_{\mu<\lambda}u_{\lambda\mu}(q_{\rm M},t_{\rm M})m_\mu$, where $\mu<\lambda$ denotes dominance ordering among partitions of the same weight. The coefficients are determined by the simultaneous eigenvalue equations of the commuting Macdonald difference operators~\cite{FeiginMacdonaldWheel2003}. The operator in Eq.~\eqref{eq:macdonald_eigenproblem} applies for arbitrary $N$.

To recover the Read--Rezayi contact state in the Jack limit, we set $q_{\rm M}=t_{\rm M}^{-(k+1)}$, corresponding to $r=2$ in the wheel specialization $t_{\rm M}^{k+1}q_{\rm M}^{r-1}=1$. This specialization can introduce poles in the rational coefficients $u_{\lambda\mu}$. For $(k,2)$-admissible partitions, however, Feigin, Jimbo, Miwa, and Mukhin proved that the specialization is well defined for generic $t_{\rm M}$ and satisfies the wheel condition~\cite{FeiginMacdonaldWheel2003}. Writing $\widehat P_\lambda(\bm x;t_{\rm M})=P_\lambda(\bm x;t_{\rm M}^{-(k+1)},t_{\rm M})$, we have
{\begin{equation}
 \widehat P_{\lambda_k^{(N)}}
 (\ldots,x,t_{\rm M}x,\ldots,t_{\rm M}^{k}x,\ldots;t_{\rm M})=0.
 \label{eq:supp_multiplicative_wheel}
\end{equation}}

For the root $\lambda=\lambda_k^{(N)}$, the specialized polynomial also has a regular limit at $t_{\rm M}=1$. In the monic normalization, the Macdonald-to-Jack limit~\cite[Sec.~II E]{JolicoeurLuque2011HighestWeight}, taken coefficientwise in the monomial basis, is
{\begin{equation}
 \lim_{t_{\rm M}\to1}\widehat P_\lambda(\bm x;t_{\rm M})
 =J_\lambda^{(-(k+1))}(\bm x).
 \label{eq:supp_regular_jack_limit}
\end{equation}}
See Sec.~\ref{sec:supp_minimal_degree} for the regularity at $t_{\rm M}=1$.

We impose this regular specialization before expanding in $\epsilon$. To express the wheel condition in particle coordinates, define $Z_{\rm cm}=N^{-1}\sum_jz_j$ and $\xi_i=z_i-Z_{\rm cm}$, and set $x_i=e^{\epsilon\xi_i}$ and $t_{\rm M}=e^{\epsilon\delta}$. If any $k+1$ particles occupy $z_{i_\ell}=z_0+\ell\delta$, with $\ell=0,\ldots,k$, then $x_{i_\ell}=Xt_{\rm M}^{\ell}$, where $X=e^{\epsilon(z_0-Z_{\rm cm})}$. By Equation~\eqref{eq:supp_multiplicative_wheel}, the resulting Taylor series therefore vanishes identically on this cluster, irrespective of the remaining particle positions.

Each Taylor coefficient satisfies the same clustering condition and is symmetric and translation invariant. The coefficient of $\epsilon^m$ is homogeneous of total degree $m$ in $(\bm z,\delta)$, since a common rescaling of $\bm z$ and $\delta$ can be absorbed into $\epsilon$. As proved in Sec.~\ref{sec:supp_minimal_degree}, the first nonzero term occurs at order $\epsilon^{D_{N,k}}$. We define the clustering polynomial through this leading term:
{\begin{equation}\color{blue}
 \widehat P_{\lambda_k^{(N)}}
 \!\left(e^{\epsilon\xi_1},\ldots,e^{\epsilon\xi_N};e^{\epsilon\delta}\right)
 =\epsilon^{D_{N,k}}\mathcal N_{N,k}^{-1}
 \mathcal W_{N,\delta}^{(k)}(\bm z)
 +O(\epsilon^{D_{N,k}+1}),
 \label{eq:supp_macdonald_leading}
\end{equation}}
where $\mathcal N_{N,k}$ is chosen so that $\mathcal W_{N,0}^{(k)}=J_{\lambda_k^{(N)}}^{(-(k+1))}$. For the monic normalization used here, $\mathcal N_{N,k}=1$, as shown in Sec.~\ref{sec:supp_minimal_degree}. With $\delta$ held fixed, coefficient extraction yields an exact polynomial satisfying Eq.~\eqref{eq:wheel} for arbitrary positions of the remaining particles. Its highest-degree part in $\bm z$ is the contact Jack polynomial, while finite $\delta$ can introduce lower-degree terms, as illustrated by Eq.~\eqref{eq:supp_n4_explicit}. The root partition and dominance ordering therefore refer to the generating Macdonald polynomial in $\bm x$, rather than to the full clustering polynomial in $\bm z$.

\begingroup

\subsection{\texorpdfstring{Minimal total degree and uniqueness}{Minimal total degree and uniqueness}}
\label{sec:supp_minimal_degree}

We first recall the characterization of symmetric polynomials satisfying contact clustering established by Feigin, Jimbo, Miwa, and Mukhin (FJMM) (Theorem~4.2 of Ref.~\cite{FeiginJackIdeal2002}).

\begingroup
\begin{theorem}[Contact-clustering space; Feigin--Jimbo--Miwa--Mukhin]
\label{thm:supp_contact_jack}

Let $k\ge1$ and $N\ge k$. The space of symmetric polynomials in $N$ complex variables that vanish whenever any $k+1$ coordinates coincide, for arbitrary values of the common coordinate and all remaining coordinates, has a basis consisting of the Jack polynomials $J_\lambda^{(-(k+1))}$ indexed by partitions $\lambda_1\ge\cdots\ge\lambda_N\ge0$ satisfying
\begin{equation}
 \lambda_i-\lambda_{i+k}\ge2,\qquad 1\le i\le N-k.
 \label{eq:supp_contact_admissibility}
\end{equation}
The Jack polynomials are taken in the monic normalization used above and are regular at $\alpha=-(k+1)$ for these admissible partitions.
\end{theorem}
\endgroup

The case $N=k$ is immediate; in the arguments below we therefore assume $N>k$.

For $N=kM$, admissibility imposes the lower bound $D_{N,k}$ on the degree of every nonzero contact-clustering polynomial. To derive this bound, iterate the admissibility inequality in steps of $k$. For a given $i$, let $q_i=\lfloor(N-i)/k\rfloor$, the largest integer for which $i+q_i k\le N$. For $q_i\ge1$, iterating admissibility and using nonnegativity of the partition parts gives
{\begin{equation}
 \lambda_i\ge\lambda_{i+k}+2\ge\cdots
 \ge\lambda_{i+q_i k}+2q_i\ge2q_i
 =2\left\lfloor\frac{N-i}{k}\right\rfloor.
 \label{eq:supp_admissibility_iteration}
\end{equation}}
When $q_i=0$, this is simply $\lambda_i\ge0$. Summing the individual lower bounds gives
{\begin{equation}
 |\lambda|\ge\sum_{i=1}^{N}2\left\lfloor\frac{N-i}{k}\right\rfloor
 =2k\sum_{j=0}^{M-1}j=kM(M-1)=D_{N,k}.
 \label{eq:supp_contact_degree_sum}
\end{equation}}
Each Jack polynomial is homogeneous of degree $|\lambda|$. By Theorem~\ref{thm:supp_contact_jack}, these admissible Jacks form a basis, so there is no nonzero contact-clustering polynomial of degree below $D_{N,k}$.

At the minimum degree $D_{N,k}$, the contact-clustering solution is unique up to normalization. Equality in the degree bound requires every $\lambda_i$ to attain its individual lower bound, since no part can decrease to compensate for an increase in another. Hence the only admissible partition of weight $D_{N,k}$ is $\lambda_k^{(N)}$ in Eq.~\eqref{eq:supp_rr_root}. By the homogeneous Jack basis in Theorem~\ref{thm:supp_contact_jack}, every solution at this degree is therefore proportional to the Read--Rezayi Jack $J_{\lambda_k^{(N)}}^{(-(k+1))}$~\cite{Bernevig_Jack08}.

The minimum-degree Jack is also invariant under center-of-mass (COM) translations, a property used below to identify the leading Macdonald coefficient. For the remainder of this subsection, write $\lambda=\lambda_k^{(N)}$ and abbreviate $J_\lambda^{(-(k+1))}$ as $J_\lambda$. Use complex particle coordinates $\bm z=(z_1,\ldots,z_N)$, and let $\bm1=(1,\ldots,1)$ and $L_-=\sum_{i=1}^N\partial_{z_i}$. A COM translation $\bm z\mapsto\bm z+a\bm1$, with $a\in\mathbb C$, shifts all particle coordinates by the same amount and preserves any coincidence of $k+1$ coordinates. Thus, on every contact configuration, $J_\lambda(\bm z+a\bm1)=0$ for all $a$; differentiating at $a=0$ shows that $L_-J_\lambda$ also satisfies contact clustering. Since $L_-$ preserves symmetry and lowers degree by one, a nonzero $L_-J_\lambda$ would contradict the minimum-degree bound. Hence $L_-J_\lambda\equiv0$. For any fixed $\bm z$, it follows that
{\begin{equation}
 \frac{d}{da}J_\lambda(\bm z+a\bm1)=(L_-J_\lambda)(\bm z+a\bm1)=0.
 \label{eq:supp_jack_translation_derivative}
\end{equation}}
Therefore $J_\lambda(\bm z+a\bm1)$ is independent of $a$ and equals its value at $a=0$, proving $J_\lambda(\bm z+a\bm1)=J_\lambda(\bm z)$ for every configuration.

The contact-state uniqueness also establishes the regular Macdonald--Jack limit in Eq.~\eqref{eq:supp_regular_jack_limit}. If a coefficient had a pole at $t_{\rm M}=1$, let $p>0$ be the largest pole order. Then $R(\bm x)=\lim_{t_{\rm M}\to1}(t_{\rm M}-1)^p\widehat P_\lambda(\bm x;t_{\rm M})$ would be a nonzero symmetric polynomial homogeneous of degree $D_{N,k}$. The wheel identity, Eq.~\eqref{eq:supp_multiplicative_wheel}, multiplied by $(t_{\rm M}-1)^p$ and taken to the limit, makes $R$ satisfy contact clustering, so $R=cJ_\lambda$. Because $\widehat P_\lambda$ is monic, the $m_\lambda$ coefficient in $R$ is $\lim_{t_{\rm M}\to1}(t_{\rm M}-1)^p=0$, whereas that in $J_\lambda$ is one. This forces $c=0$, a contradiction. The finite limit itself retains contact clustering and the unit $m_\lambda$ coefficient, so it equals $J_\lambda$. This establishes Eq.~\eqref{eq:supp_regular_jack_limit} and justifies the Taylor expansion used below.

We now establish the corresponding minimum-degree and uniqueness result for the finite-displacement clustering polynomials underlying our lattice Read--Rezayi states. The proof uses Theorem~\ref{thm:supp_contact_jack}, the Macdonald construction in Sec.~\ref{sec:supp_general_macdonald}, and the contact-clustering condition inherited by the lowest nonzero coefficient in the displacement expansion.

\begin{samepage}
\begingroup
\color{blue}\begin{theorem}[Minimum degree and uniqueness at finite displacement]
\label{thm:supp_displaced_uniqueness}

Let $N=kM$, with positive integers $k,M$, and fix a nonzero complex displacement $\delta_0$. Consider nonzero symmetric, translation-invariant polynomials $f(\bm z)$ satisfying
\begin{equation}
 f(\ldots,z_0,z_0+\delta_0,\ldots,z_0+k\delta_0,\ldots)=0
 \label{eq:supp_displaced_clustering_condition}
\end{equation}
for every choice of $k+1$ particle labels, every complex $z_0$, and arbitrary remaining coordinates. Their minimum total coordinate degree is $D_{N,k}=N(N-k)/k$, attained by the Macdonald-derived polynomial $\mathcal W_{N,\delta_0}^{(k)}$ in Eq.~\eqref{eq:supp_macdonald_leading}.
Every solution at this degree is proportional to $\mathcal W_{N,\delta_0}^{(k)}$; the minimum-degree solution is therefore unique up to an overall constant within this class.
\end{theorem}
\endgroup
\end{samepage}

The physical meaning of Theorem~\ref{thm:supp_contact_jack} and Theorem~\ref{thm:supp_displaced_uniqueness} is as follows. Theorem~\ref{thm:supp_contact_jack} implies that the continuum Read--Rezayi model wavefunction constructed from the Jack polynomial is the unique minimum-degree zero mode of the $(k+1)$-body contact parent Hamiltonian, with total polynomial degree $D_{N,k}$. At higher total degrees, additional zero modes are allowed, including quasihole states; below $D_{N,k}$, no zero mode exists. Theorem~\ref{thm:supp_displaced_uniqueness} yields a similar result for lattice clustering model states. Within the stated class of symmetric, COM-translation-invariant planar polynomials, the minimum total coordinate degree remains $D_{N,k}$, and the polynomial satisfying finite-displacement clustering at this degree is unique up to normalization.

The proof has four steps: a degree bound for jointly homogeneous families, existence by the Macdonald construction, uniqueness from the FJMM contact result, and homogenization to cover arbitrary candidates at fixed displacement. In Steps~1--3, total degree counts both $\bm z$ and $\delta$; Step~4 concerns coordinate degree at fixed $\delta_0$.

\begin{proof}
\emph{Step 1: Degree bound for homogeneous families.---}Let $F_m(\bm z,\delta)\not\equiv0$ be symmetric and COM translation invariant in $\bm z$, homogeneous of total degree $m$ in $(\bm z,\delta)$, and satisfy finite-displacement clustering for every $\delta$. Expand in the independent variable $\delta$:
{\begin{equation}
 F_m(\bm z,\delta)=\sum_{j=s}^{m}\delta^jA_j(\bm z),
 \qquad A_s\not\equiv0,
 \label{eq:supp_homogeneous_delta_expansion}
\end{equation}}
where $s$ is the smallest index with $A_s\not\equiv0$: $s=0$ if $A_0\not\equiv0$, and $s>0$ otherwise. The coefficient $A_s$ is homogeneous of degree $m-s$ and inherits permutation symmetry and COM translation invariance by comparing coefficients of $\delta$ in the corresponding identities for $F_m$.

The lowest nonzero coefficient $A_s$ satisfies contact clustering. Substitute a displaced cluster into $F_m$ and cancel the common factor $\delta^s$ in the polynomial identity:
{\begin{equation}
 \sum_{j=s}^{m}\delta^{j-s}
 A_j(z_0,z_0+\delta,\ldots,z_0+k\delta,z_{k+2},\ldots,z_N)=0.
 \label{eq:supp_lowest_coefficient_wheel}
\end{equation}}
Setting $\delta=0$ removes every term with $j>s$ and makes the first $k+1$ arguments coincide, leaving
{\begin{equation}
 A_s(\underbrace{z_0,\ldots,z_0}_{k+1},z_{k+2},\ldots,z_N)=0.
 \label{eq:supp_lowest_coefficient_contact}
\end{equation}}
By symmetry, this holds for any $k+1$ particle labels. The contact-state degree bound derived from the FJMM theorem, Theorem~\ref{thm:supp_contact_jack}, therefore gives
{\begin{equation}
 m-s=\deg A_s\ge D_{N,k},\qquad\text{hence}\qquad m\ge D_{N,k}.
 \label{eq:supp_homogeneous_degree_bound}
\end{equation}}
The above result shows that, if there is a nonzero homogeneous polynomial obeying the lattice clustering rule, its total degree in $(\bm z, \delta)$ has to be at least $D_{N,k}$. In the next two steps, we show that a polynomial whose degree is right at $D_{N,k}$ exists and, more importantly, is unique up to normalization within this homogeneous class.

\emph{Step 2: Existence at the minimum degree.---}We now show that the Macdonald construction in Sec.~\ref{sec:supp_general_macdonald} attains the lower bound $D_{N,k}$. Write its Taylor expansion as
{\begin{equation}
 \widehat P_{\lambda}(e^{\epsilon\bm\xi};e^{\epsilon\delta})
 =\sum_{m\ge0}\epsilon^m C_m(\bm z,\delta).
 \label{eq:supp_macdonald_taylor_series}
\end{equation}}
Each $C_m$ is a polynomial, homogeneous of total degree $m$ in $(\bm z,\delta)$, with the symmetry and clustering properties required in Step~1, so $C_m=0$ for $m<D_{N,k}$. To show that $C_{D_{N,k}}$ is nonzero, use the regular Jack limit, Eq.~\eqref{eq:supp_regular_jack_limit}, at $\delta=0$. COM translation invariance and degree-$D_{N,k}$ homogeneity of $J_\lambda$ give
{\begin{equation}
\begin{aligned}
 \sum_{m\ge0}\epsilon^m C_m(\bm z,0)
 &=J_\lambda(e^{\epsilon\bm\xi})
 =J_\lambda(e^{\epsilon\bm\xi}-\bm1)\\
 &=\epsilon^{D_{N,k}}J_\lambda(\bm z)
   +O(\epsilon^{D_{N,k}+1}),
\end{aligned}
 \label{eq:supp_contact_leading}
\end{equation}}
where $e^{\epsilon\xi_i}-1=\epsilon\xi_i+O(\epsilon^2)$ and $J_\lambda(\bm\xi)=J_\lambda(\bm z)$. Comparing coefficients yields $C_{D_{N,k}}(\bm z,0)=J_\lambda(\bm z)\not\equiv0$. Thus $\mathcal W_{N,\delta}^{(k)}=C_{D_{N,k}}$ and $\mathcal N_{N,k}=1$ in Eq.~\eqref{eq:supp_macdonald_leading}. Joint homogeneity makes every term with a positive power of $\delta$ have lower coordinate degree, so at any fixed $\delta$ the constructed polynomial is nonzero and has coordinate degree exactly $D_{N,k}$.

\emph{Step 3: Uniqueness of the homogeneous family.---}Every homogeneous solution at degree $D_{N,k}$ is proportional to the one constructed in Step~2. Let $F_{D_{N,k}}(\bm z,\delta)\not\equiv0$ be homogeneous of total degree $D_{N,k}$ in $(\bm z,\delta)$ and satisfy the symmetry and clustering conditions of Step~1. If its lowest nonzero power of $\delta$ is $s$, Eq.~\eqref{eq:supp_homogeneous_degree_bound} gives $D_{N,k}-s\ge D_{N,k}$, hence $s=0$. Its contact value is therefore nonzero. By the one-dimensionality of the minimum-degree contact-clustering space derived from the FJMM theorem, Theorem~\ref{thm:supp_contact_jack}, it follows that $F_{D_{N,k}}(\bm z,0)=cJ_\lambda(\bm z)$ with $c\ne0$.

The difference $F_{D_{N,k}}(\bm z,\delta)-c\mathcal W_{N,\delta}^{(k)}(\bm z)$ satisfies the same symmetry, homogeneity, and clustering conditions, and vanishes identically at $\delta=0$. If nonzero, its lowest power of $\delta$ would therefore be positive, contradicting $s=0$. Hence $F_{D_{N,k}}(\bm z,\delta)=c\mathcal W_{N,\delta}^{(k)}(\bm z)$, proving uniqueness up to normalization.

\emph{Step 4: Fixed displacement.---}We now pass from homogeneous families to arbitrary polynomials at fixed $\delta_0\ne0$. Since Step~2 already supplies a degree-$D_{N,k}$ solution, it suffices to consider a nonzero symmetric, COM-translation-invariant polynomial $f(\bm z)$ satisfying clustering at $\delta_0$, with $\deg f\le D_{N,k}$. Decompose $f(\bm z)=\sum_{d=0}^{D_{N,k}}f_d(\bm z)$, where $f_d$ is its part of total coordinate degree $d$. With $\delta$ an independent polynomial variable, define
{\begin{equation}
 \widetilde f(\bm z,\delta)
 =\sum_{d=0}^{D_{N,k}}
 \left(\frac{\delta}{\delta_0}\right)^{D_{N,k}-d}f_d(\bm z).
 \label{eq:supp_fixed_delta_scaling}
\end{equation}}

The factor $(\delta/\delta_0)^{D_{N,k}-d}$ gives each term joint degree $D_{N,k}$ and equals one at $\delta=\delta_0$. Thus $\widetilde f$ is a nonzero polynomial homogeneous of total degree $D_{N,k}$ in $(\bm z,\delta)$, and $\widetilde f(\bm z,\delta_0)=f(\bm z)$. To verify its clustering and symmetries, write, for $\delta\ne0$,
{\begin{equation}
 \widetilde f(\bm z,\delta)
 =\left(\frac{\delta}{\delta_0}\right)^{D_{N,k}}
 f\!\left(\frac{\delta_0}{\delta}\bm z\right),
 \qquad\delta\ne0.
 \label{eq:supp_fixed_delta_rescaling}
\end{equation}}
Rescaling maps a cluster $z_{i_\ell}=z_0+\ell\delta$, $\ell=0,\ldots,k$, to $\delta_0z_0/\delta+\ell\delta_0$, where $f$ vanishes. Symmetry and COM translation invariance are inherited from $f$ by the same rescaling. Polynomiality of $\widetilde f$ extends these clustering and symmetry identities to $\delta=0$, so $\widetilde f$ satisfies the conditions of Step~3.

Step~3 gives $\widetilde f(\bm z,\delta)=c\mathcal W_{N,\delta}^{(k)}(\bm z)$ for some constant $c\ne0$. Setting $\delta=\delta_0$ yields $f(\bm z)=c\mathcal W_{N,\delta_0}^{(k)}(\bm z)$, whose coordinate degree is exactly $D_{N,k}$ by Step~2. Thus no lower-degree solution exists, and the solution at degree $D_{N,k}$ is unique up to normalization.
\end{proof}

\endgroup

\subsection{Abelian limit and exchange statistics}
For $k=1$, the clustering polynomial reduces to the pair factor underlying the generalized Laughlin state:
\begin{equation}
    \mathcal W_{N,\delta}^{(1)}(\bm z)
    \propto\prod_{i<j}\left[(z_i-z_j)^2-\delta^2\right].
    \label{eq:supp_k1_wheel}
\end{equation}
The corresponding $k=1$ Macdonald polynomial has the explicit product form $\widehat P_{2\rho}(\bm x;t_{\rm M})=\prod_{i<j}(x_i-t_{\rm M}x_j)(x_i-t_{\rm M}^{-1}x_j)$, with $\rho=(N-1,\ldots,0)$.  Under $x_i=e^{\epsilon\xi_i}$ and $t_{\rm M}=e^{\epsilon\delta}$, its first nonzero term occurs at order $\epsilon^{N(N-1)}$ and yields the pair product in Eq.~\eqref{eq:supp_k1_wheel}.  The pair product has degree $N(N-1)=D_{N,1}$ and reduces at $\delta=0$ to the bosonic $\nu=1/2$ Laughlin Jack polynomial.

For an inversion-symmetric displacement multiset $\{\delta_l\}_{l=1}^{m}=\{-\delta,\delta,\delta'_1,\ldots,\delta'_{m-2}\}$, choosing $p=m-2$ in Eq.~\eqref{eq:jrr_dressed} reproduces the full planar pair factor $\prod_{l=1}^{m}\prod_{i<j}(z_i-z_j+\delta_l)$ and gives $\nu=1/m$.

To illustrate the dressing, let $B_\delta$ denote the product in Eq.~\eqref{eq:supp_k1_wheel} and let $V=\prod_{i<j}(z_i-z_j)$ be the Vandermonde determinant. The planar polynomial parts of the deformed Laughlin states at $\nu=1/3$ and $1/4$ are, respectively, $B_{\delta_1}V$ and $B_{\delta_1}B_{\delta_2}$. Their dressing factors are $\mathcal J^{(1)}=V$ and $\mathcal J^{(2)}=B_{\delta_2}$; the latter corresponds to $\{\delta'_a\}=\{-\delta_2,\delta_2\}$. Taking $\delta_2=0$ gives $B_{\delta_1}V^2$, with the double contact zero used in Sec.~\ref{sec:supp_boson_laughlin14}. At $\delta_1=\delta_2=0$, this becomes $V^4$, itself the $(1,4)$ Jack state with $\alpha=-2/3$ and root $100010001\cdots$~\cite{Bernevig_Jack08}. For independent finite displacements, these states are constructed as products of pair factors; a single $(1,4)$ Macdonald representation is not assumed. Here ``Macdonald-derived'' refers specifically to $\mathcal W_{N,\delta}^{(k)}$ in Eq.~\eqref{eq:jrr_dressed}.

For general $k$, the inversion-symmetric displaced Jastrow factor in Eq.~\eqref{eq:jrr_dressed} preserves the $(k+1)$-particle clustering zero.  It contributes the exchange sign $(-1)^p$, giving bosonic statistics for even $p$ and fermionic statistics for odd $p$. When all additional displacements vanish, it reduces to $\prod_{i<j}(z_i-z_j)^p$.

\FloatBarrier

\section{Extension to Jain states}
\label{sec:supp_jain_multiband}
We now explore whether the finite-displacement construction can be extended to Jain states. For conventional Jain states, projection to the lowest Landau level obscures the explicit Jastrow zeros on which our construction relies~\cite{jain2007composite}. We therefore consider the unprojected $\nu=2/5$ Jain state, which lies within the lowest two Landau levels, retains explicit Jastrow zeros, and admits a local parent Hamiltonian~\cite{RezayiMacDonald1991Jain,BandyopadhyayJainParent2020}.

We first use torus exact diagonalization of the continuum parent Hamiltonian to establish a finite-size reference. We then introduce finite displacements into the planar Jastrow factor and construct a lattice kinetic Hamiltonian with two zero-energy bands spanning the sampled $0$LL and $1$LL orbitals. Finally, we test whether the corresponding finite-displacement density interaction isolates a five-state zero-energy manifold within this two-band space on finite tori. We find more than five zero modes, indicating that additional constraints are needed to isolate the Jain manifold.

\subsection{Continuum reference: wavefunction, parent Hamiltonian, and torus spectrum}

Let $\Phi_n$ denote the Slater determinant for $n$ filled effective Landau levels.  On the plane, the unprojected Jain state at $\nu=n/(2n+1)$ takes the form
{\begin{equation}
 \Psi_{n/(2n+1),0}^{\rm J}(\bm z,\bar{\bm z})
 =\Phi_n(\bm z,\bar{\bm z})
 \prod_{i<j}(z_i-z_j)^2,
 \label{eq:supp_unprojected_jain}
\end{equation}}
where the Gaussian factor is omitted.
Fermionic antisymmetry ensures that $\Phi_n$ vanishes at least linearly as two particles approach each other, while the Jastrow factor contributes a quadratic zero. The resulting zero of at least third order makes the gradient with respect to the relative coordinate $\bm r=\bm r_i-\bm r_j$ vanish when the two particles coincide. The Trugman--Kivelson interaction $V_{\rm TK}(\bm r)=(g_{\rm TK}/2)\nabla^2\delta^{(2)}(\bm r)$, with $g_{\rm TK}>0$, penalizes nonzero relative-coordinate gradients at particle coincidence for fermionic states~\cite{TrugmanKivelson85,RezayiMacDonald1991Jain}. Since these gradients vanish for the state in Eq.~\eqref{eq:supp_unprojected_jain}, it is an exact zero mode of this interaction. Within the LLL, this interaction is proportional to Haldane's fermionic $V_1$ pseudopotential.  Restricted to the lowest two Landau levels, this interaction gives the Rezayi--MacDonald parent Hamiltonian for the unprojected $\nu=2/5$ Jain state, which serves as our continuum reference.

We extend the magnetic-Bloch exact-diagonalization framework used above to the continuum $0$LL+$1$LL subspace, treating the two Landau levels as degenerate~\cite{wang_exact_2021,Liu_2025}. The interaction matrix elements, including those coupling the two Landau levels, are evaluated using the explicit form-factor expressions in Appendix~E of Ref.~\cite{Liu_2025}. Figure~\ref{fig:supp_jain_continuum_torus} shows the resulting finite-size spectrum.

\begin{figure}[htbp]
 \centering
 \includegraphics[width=2.15in]{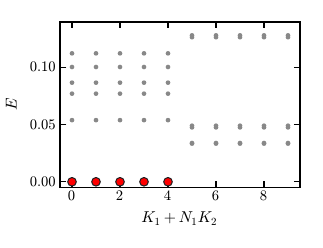}
 \caption{\textbf{Momentum-resolved torus spectrum of the continuum $\nu=2/5$ parent.}
 The six lowest energies per momentum sector are shown for the Trugman--Kivelson interaction in the degenerate $0$LL+$1$LL subspace, with $N_e=4$, $N_\phi=10$, $N_1\times N_2=5\times2$ magnetic unit cells, and $L_x/L_y=5/2$. The five zero modes (red) are separated from the excited states (gray) by a finite-size gap $\Delta=0.03343$. We set $\ell_B=1$ and $g_{\rm TK}=8\pi$.}
 \label{fig:supp_jain_continuum_torus}
\end{figure}

\subsection{Finite-displacement wavefunction and lattice density interaction}

On the plane, we displace the two Jastrow zeros to $z_i-z_j=\pm\delta$, leaving the Slater determinant $\Phi_n$ unchanged:
{\begin{equation}\color{blue}
 \Psi_{n/(2n+1),\delta}^{\rm J}(\bm z,\bar{\bm z})
 =\Phi_n(\bm z,\bar{\bm z})
 \prod_{i<j}\left[(z_i-z_j)^2-\delta^2\right].
 \label{eq:supp_displaced_jain}
\end{equation}}
At $\delta=0$, this state reduces to the unprojected Jain state in Eq.~\eqref{eq:supp_unprojected_jain}.  The displaced Jastrow factor remains symmetric under particle exchange, so the state retains the fermionic antisymmetry of $\Phi_n$.  This Jastrow factor is the $k=1$ Macdonald-derived clustering polynomial given in Eq.~\eqref{eq:supp_k1_wheel}.  The deformation preserves the highest power of each coordinate in the Jastrow factor, leaving the planar flux counting unchanged. It also leaves the antiholomorphic dependence unchanged, so the state remains within the lowest $n$ Landau levels.

When $\bm\delta$ is a lattice vector, restricting the particle coordinates to lattice sites makes the wave function vanish for every configuration in which both $\bm R$ and $\bm R+\bm\delta$ are occupied.  The lattice state is therefore an exact zero mode of the positive-semidefinite density interaction
{\begin{equation}\color{blue}
 \widehat H_{\bm\delta}^{\rm J}
 =U\sum_{\bm R}\hat n_{\bm R}\hat n_{\bm R+\bm\delta},
 \qquad U>0,
 \label{eq:supp_displaced_jain_parent}
\end{equation}}
without requiring band projection. To test whether this density constraint isolates the Jain manifold on a finite torus, we impose it in the lattice subspace spanned by sampled $0$LL and $1$LL orbitals satisfying magnetic boundary conditions. The zero-mode count can be obtained directly from the projected interaction, without constructing an explicit torus counterpart of Eq.~\eqref{eq:supp_displaced_jain}.

\subsection{Two-band lattice realization and zero-mode test}

We construct a lattice kinetic Hamiltonian whose zero-energy subspace is spanned by sampled $0$LL and $1$LL orbitals.  Following Cruise and Seidel~\cite{Seidel_NonAbelian_26}, we sample the torus $0$LL and $1$LL orbitals at the lattice sites within one magnetic unit cell for each magnetic Bloch momentum $\bm k$.  The resulting lattice Bloch vectors, $|u_0(\bm k)\rangle$ and $|u_1(\bm k)\rangle$, form the columns of $\mathsf V(\bm k)=\bigl(|u_0(\bm k)\rangle,|u_1(\bm k)\rangle\bigr)$.  Since lattice sampling need not preserve orthonormality, we introduce the $2\times2$ Gram matrix $G(\bm k)=\mathsf V^\dagger(\bm k)\mathsf V(\bm k)$.  When $\det G(\bm k)>0$, the two sampled orbitals are linearly independent, and $\Pi_{2{\rm LL}}(\bm k)=\mathsf V(\bm k)G^{-1}(\bm k)\mathsf V^\dagger(\bm k)$ projects the full lattice Bloch space onto their two-dimensional span.  For the equal-weight choice of Ref.~\cite{Seidel_NonAbelian_26}, the positive-semidefinite Bloch Hamiltonian is obtained by multiplying the complementary projector $I-\Pi_{2{\rm LL}}(\bm k)$ by $\det G(\bm k)$. 
{\begin{equation}
 h_{2{\rm LL}}(\bm k)
 =\det G(\bm k)\left[I-\Pi_{2{\rm LL}}(\bm k)\right],\qquad \det G(\bm k)>0.
 \label{eq:supp_exact_two_ll}
\end{equation}}

\begin{figure}[H]
 \centering
 \includegraphics[width=2.15in]{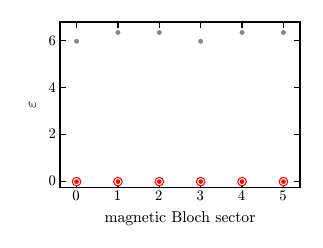}
 \caption{\textbf{Exact two-Landau-level lattice kinetic benchmark.}
 This single-particle example uses an $M\times N=4\times6$ lattice with flux density $\phi=1/(mn)=1/4$, an $m\times n=2\times2$ magnetic unit cell, and $N_\phi=6$.  Two coincident zero-energy eigenvalues (red) occur in every sector and span the sampled $0$LL+$1$LL subspace; positive bands are gray.}
 \label{fig:supp_jain_lattice}
\end{figure}

The sampled orbitals obey $h_{2{\rm LL}}(\bm k)|u_j(\bm k)\rangle=0$ for $j=0,1$, while every orthogonal state has energy $\det G(\bm k)$. The model therefore has exactly two zero-energy flat bands, separated from the remaining bands whenever $\det G(\bm k)>0$ throughout the magnetic Brillouin zone. Figure~\ref{fig:supp_jain_lattice} illustrates this construction for a $4\bm a_1\times6\bm a_2$ lattice with $N_\phi=6$; we use it also for the larger tori below.

To test whether the density constraint selects the Jain manifold, we project the density interaction in Eq.~\eqref{eq:supp_displaced_jain_parent}, with $U=1$ and $\bm\delta=\bm a_1$, into the sampled $0$LL+$1$LL subspace. Both the kinetic benchmark and the interaction calculations below use $(\theta_x,\theta_y)=(0,0)$. \textcolor{blue}{For $N_e=4$ and $N_\phi=10$ on a $10\bm a_1\times4\bm a_2$ lattice at $\phi=1/4$, the projected interaction has $175$ zero modes, far more than the five of the continuum Jain parent. For a larger system with $N_e=6$ and $N_\phi=15$ on a $10\bm a_1\times6\bm a_2$ lattice at $\phi=1/4$, the ten lowest states in each of the $15$ momentum sectors have numerically zero energy, establishing a lower bound of $150$ zero modes. These results show that realizing the required two-band subspace and imposing the displaced pair zeros are not sufficient to isolate the Jain manifold. Additional interaction constraints are needed.}

\end{document}